# Resolving room temperature microscale fracture and plasticity of iron oxides along the cascade of iron ore reduction via nanoindentation and microcantilever bending

**Shreehard Sahu*, James P. Best, Gerhard Dehm and Anwesha Kanjilal***
*Max Planck Institute for Sustainable Materials, 40237 Düsseldorf, Germany*

**Corresponding authors: s.sahu@mpi-susmat.de, a.kanjilal@mpi-susmat.de*

**Abstract**

Understanding the fundamental mechanical behaviour of iron oxide phases is essential for controlling attrition and fracture during iron ore reduction process, particularly in hydrogen-based direct reduction systems. This study investigates the room temperature plasticity and fracture behaviour of single-crystal hematite, magnetite, and Wüstite using nanoindentation and micro-cantilever fracture testing. Hematite exhibited the highest hardness, ***H*** and elastic modulus, ***E*** (***H***=18.5±0.5 GPa, ***E***=281±3 GPa), followed by magnetite (***H***=8.7±0.5 GPa, ***E***=165±3 GPa) and Wüstite (***H***=7.5±0.4 GPa, ***E***=145±4 GPa), reflecting differences in slip activity along the iron oxide reduction sequence. Furthermore, fracture toughness was measured using notched microcantilevers for all three iron oxide phases, aligned along low index and high index crystallographic planes, respectively. For the low index-oriented case hematite showed increased fracture toughness owing to crack deviation and faceting while magnetite and Wüstite exhibited single plane cleavage fracture. Distinct changes in the deformation behavior in terms of plasticity and cracking of the three iron oxides were evident from both methods. Further investigation of a magnetite–gangue interface, particularly relevant to low-concentration ores, revealed significantly reduced fracture toughness compared to the magnetite phase. Overall, these results provide a comprehensive set of mechanical properties of iron oxides with potential application in material models for predicting fracture and attrition during hydrogen-based direct reduction.

## 1. Introduction

Iron ore pellets have long been the preferred feedstock for iron making through direct reduction routes, primarily due to their superior mechanical strength as compared to lump ore or sintered feed. However, these pellets are also prone to surface abrasion during various stages of material handling, leading to the generation of fines. In a conventional blast furnace, the mechanical properties of the iron oxide feedstock are designed to avoid reduction disintegration at the upper blast [1,2]. However, in a hydrogen-based direct reduction (HyDR) system, the reducing gas flow rates are much higher compared to conventional blast furnaces [3,4], which results in a higher degree of mechanical stress on the cold feedstock. These mechanical stresses caused by inter-pellet and pellet wall collision result in attrition, leading to the formation of iron ore fines, which are lost along with the reducing gas flow [1–3]. This loss of fines is significant at room temperature [5]. Furthermore, during solid-state reduction, controlled fracture of the partially reduced feedstock can accelerate the reaction in its final stages (albeit at high temperatures), which is otherwise typically sluggish [6,7]. Hence, the increased susceptibility to mechanical degradation, combined with the need for controlled fracture during solid-state reduction, highlights the importance of optimizing the mechanical properties of the feedstock. Previous studies of attrition in iron oxide pellets is primarily limited to the role of extrinsic properties such as pellet size and velocity [2,3,8]. However, there has been limited research into the intrinsic mechanical properties of feedstock materials, specifically hematite and magnetite.

Attrition can be classified into two categories: "surface abrasion" and "fragmentation" [9–11]. Ghadiri [12,13], using an indentation-based mechanistic model, demonstrated the dependence of surface wear and chipping propensity on mechanical properties such as hardness ***(H)*** and fracture toughness ***($K_c$)***. In the case of fragmentation as well, the force required for crack propagation is also influenced by ***H*** and ***$K_c$*** [12]. From a computational point of view, these mechanical properties are important inputs to various large-scale material models used to simulate breakage damage of the feed stock [8,12]. It is important to note that although all these breakage mechanisms can be analysed within the framework of indentation fracture, the scaling relationships between the mechanical properties (***H*** and ***$K_c$***) and the intensity of attrition vary across mechanisms. Quinn and Quinn [14] introduced a more comprehensive parameter, the brittleness index (***Q***), which captures the balance between deformation and fracture processes by relating the energy required for plastic deformation to the energy required for crack propagation . In the present study, this parameter is used to compare the attrition behaviour of the three iron oxide phases.

The aforementioned discussions suggest that the intrinsic material properties which govern the attrition behavior of the feedstock are hardness and fracture toughness, which in turn depend on dislocation activity and crack propagation trajectories. Hence, a fundamental understanding of the deformation and fracture mechanisms of iron oxide phases becomes essential. However, commercial bulk iron ore pellets exhibit microstructural complexity, containing porosity, microcracks and gangue phases which complicates the understanding of the deformation mechanisms and related measurement of mechanical properties such as $\boldsymbol{H}$ and $\boldsymbol{K_c}$ of the iron oxides [6,15,16]. Previous studies [17,18] have often relied on iron oxide samples prepared either through the oxidation of metallic iron or by sectioning from natural ores, both of which inherently possess the aforementioned microstructural heterogeneities. Furthermore, due to the intrinsically brittle nature of iron oxides, most prior investigations have been restricted to high temperature creep experiments to elucidate deformation mechanisms [19–23], leaving a knowledge gap in the room temperature deformation behavior which is critical for understanding the attrition of iron oxide ores.

Understanding plastic flow at room temperature in such macroscopically brittle oxides, requires suppressing fracture before plasticity which is challenging in the absence of any confining pressure [17,24–27]. Kollenberg [28] used natural single crystal hematite and reported room-temperature Vickers hardness values for hematite ranging from 10 to 14 GPa, depending on the crystallographic orientation . In contrast, room-temperature critical resolved shear stress (CRSS) measurements by Siemes for various slip systems were an order of magnitude lower (~90–500 MPa) [17]. This disparity is likely caused due to the use of macroscopic single crystals derived from natural ore, by Siemes, which inherently contain microstructural heterogeneities as discussed previously [17]. Likewise, a systematic assessment of fracture behavior of the iron oxide phases is also lacking. Madadi [8] reported a fracture toughness of 2.3 ± 1 MPa$\sqrt{m}$ for hematite, using Warren's indentation-based technique [29], on a commercial iron ore pellet with significant microstructural variability. For magnetite and Wüstite, both room-temperature plasticity and fracture behavior remain essentially unexplored, underscoring a critical gap in understanding mechanical response across iron oxide phases in the reduction cascade. These challenges can be circumvented by small-scale testing strategies, such as nanoindentation and microcantilever fracture testing, which require confined testing of small volume, thereby minimizing artifacts from pre-existing flaws [30–34]. The use of single crystals further enables elucidating and quantifying orientation specific deformation processes and fracture toughness values of targeted crystallographic planes and the corresponding crack trajectories, which are discussed later. In practice, however, industrial iron ore pellets also contain gangue particles often up to 7-15 volume % in low grade ores [35,36] underscoring the need to elucidate the role of gangue-iron oxide interface fracture on attrition properties.

In this context, the present study aims to characterize the room-temperature plasticity mechanisms and quantify the fracture toughness of hematite, magnetite, and Wüstite using small scale mechanical testing on model single crystals free from preexisting flaws and porosities. In addition, fracture along the gangue-iron oxide interface was investigated. Nanoindentation and micro-cantilever fracture testing are employed to determine indentation modulus $\boldsymbol{E}$, hardness $\boldsymbol{H}$ and fracture toughness $\boldsymbol{K_c}$, respectively—mechanical parameters that are essential for evaluating the chipping and cracking tendencies of individual iron oxide phases through the brittleness index $\boldsymbol{Q}$.

## 2. Experimental Procedure

### *2.1. Materials*

Single crystals of iron oxide phases - hematite, magnetite and Wüstite were procured from Mateck Materials -Technologies & Kristalle GmbH (Germany) with the surface planes mentioned in Table 1. The single crystals of hematite and magnetite were cut from their respective natural ores, whereas Wüstite was produced using the Czochralski method. The orientation of the as-received single crystals was initially confirmed using X-Ray diffraction (XRD), performed using Rigaku XRD machine using Bragg-Brentano method. The stoichiometry of Wüstite was checked using Electron Probe Microanalyzer (EPMA), and the corresponding details are mentioned in Supplementary Information Section 1.

**Table 1**: Surface plane orientation of iron oxide single crystals

| Iron Oxide phases | Surface plane orientation |
|---|---|
| Hematite | $(0001), (0\bar{1}10)$ & $(1\bar{1}02)$ |
| Magnetite | $(001)$ |
| Wüstite | $(001)$ |

### *2.2. Mechanical testing and post-mortem characterization*

Nanoindentation was carried out using G200 Nanoindenter (Keysight Technologies, Santa Rose, CA, USA) at ambient temperature (21$^{\circ}$ C), on basal $(0001)$ orientation for hematite and cubic $(001)$ plane for magnetite and Wüstite. A Berkovich diamond indenter was used to measure the indentation modulus ($\boldsymbol{E}$) and hardness ($\boldsymbol{H}$) using the continuous stiffness measurement (CSM) technique [37]. The $\boldsymbol{E}$ and $\boldsymbol{H}$ values were averaged over indentation depths between 150-250 nm. Over 25 indents were performed on each sample to ensure reproducibility. Additionally, spherical indentation was performed on the same crystallographic planes mentioned previously, as well as on the $(0\bar{1}10)$ and $(1\bar{1}02)$ planes of hematite, using a 1 μm-diameter diamond tip to investigate the active deformation mechanisms and crack formation associated with the observed surface features. In both cases, loading was done at constant indentation strain rate of 0.05 $s^{-1}$ up to a depth of 300 nm. Additionally, in magnetite and Wüstite the spherical indentation was also done to higher depths of 1500 nm to understand the cracking behavior and pile up formation around the indents. The orientation of the slip steps and cracks around the spherical indent was characterized using a Zeiss Merlin SEM, in conjunction with EBSD ensuring no change in the in-plane orientation. This was achieved by using an FIB marker as reference during EBSD and SEM imaging. The SEM images of the indents and the corresponding crystallographic data from EBSD were used as input into a Matlab® code which can identify the surface traces from different slip systems [38].

Fracture toughness measurements were performed using simple rectangular and pentagonal microcantilever geometries, shown in Fig. 1a,b, across two distinct sets of crystallographic planes in the three iron oxide phases. In the first set, the notches were aligned along low-index planes, which are known to be preferred cleavage planes [39]. The specific orientations of these notch planes are provided in Table 2. To determine the upper bound of fracture toughness, additional tests were conducted with the notch oriented along random high-index planes for the iron oxide phases, also listed in Table 2. The magnetite single crystal contained a few widely separated gangue particles of 50-100 μm size, some of which exhibited a well-defined straight interface with the magnetite phase. To further understand the fracture along this gauge particle-iron oxide interface, microcantilevers were also fabricated at the interface between gangue particle-magnetite. The details of the chemical analysis revealed Si, Mg, Al and O as the main constituents of the gangue particle as shown in Fig. S1 of Supplementary Information. Depending on the orientation of the notch plane and ease of fabrication, some cantilevers were fabricated with rectangular cross-sections (as shown in Fig. 1a), while others were fabricated with pentagonal cross section (as shown in Fig. 1b). This implies rectangular beams were employed when fabrication along the sample edge was feasible, whereas pentagonal cross-sections were adopted for orientations requiring fabrication away from the edge. The dimensions of the cantilevers were selected such that plane strain condition is valid, which requires the thickness ***(B)*** (as annotated in Fig. 1a) of the cantilever to be ~25 times the radius of the plastic zone ***(r)***, as given in Eq. 1 & 2 [40].

**Table 2**: List of low and high index orientations and cross section geometries used for microcantilever fracture test of hematite, magnetite and Wüstite.

| Material | Low Index Orientations | | High index orientation | |
|---|---|---|---|---|
| | Notch plane | Cross sectional shape | Notch plane | Cross sectional shape |
| **Hematite** | (0001) | Rectangular | $(\overline{14}\ 7\ 7\ 30)$ | Rectangular |
| **Magnetite** | (001) | Pentagon | Near (012) | Pentagon |
| **Wüstite** | (001) | Rectangular | Near (013) | Pentagon |

$$\boldsymbol{r} = \frac{K_{IC}^2}{3\pi\sigma_Y^2} \quad (1)$$

$$\boldsymbol{B} \geq 2.5\ \frac{K_{IC}^2}{\sigma_Y^2} \quad (2)$$

where $\boldsymbol{K_{IC}}$ is the Mode I fracture toughness and $\boldsymbol{\sigma_y}$ is the yield strength of the iron oxide phases (obtained from nanoindentation in the present study as shown in Fig. S2 of Supplemental Information). Using the above formulation, the sample thickness was fixed based on the softest phase, i.e. Wüstite [11,15].

The microcantilevers were fabricated using $Xe^+$ plasma-based focused ion beam (P-FIB) machining in a FEI Helios® G3 Cx dual beam FIB-SEM system (ThermoFisher Scientific). The $Xe^+$ PFIB was specifically selected to avoid any $Ga^+$ induced damage in hematite, which has been reported previously [41]. The coarse cutting of the microcantilevers was done using 30 kV and 2 nA current followed by fine cut at 30 kV and 100 pA. A bridge notch has been used to generate a sharp pre-crack and avoid any FIB-induced artifacts, such as rounding of notch root radius and curved notch front caused by over-fibbing which can artificially influence the fracture toughness values [42]. The dimensions of the bridge notch are compliant with Ref [42]. Final milling of the bridge notch was achieved using a line scan applying 30 pA ion beam current for a targeted notch depth ***(a)*** equating to ***a/W*** ≈ 0.2 (for rectangular microcantilevers) and ***a/2c*** ≈ 0.2 (for pentagonal microcantilevers) where ***W*** is the thickness of the rectangular microcantilever and ***c*** is the distance between the top surface and neutral plane of the pentagonal beam which is calculated from Ref [43]. For the gangue-magnetite interface system, the bridge notch was carefully aligned along the interface which was aligned to the {120} plane of magnetite, with the gangue element on the fixed end and magnetite on the loading end of the cantilever. The *in-situ* testing was carried out using Alemnis ASA (Switzerland) nanoindentation setup, mounted in a Zeiss Gemini 500 SEM using a diamond spherical indenter tip of 2 μm diameter (Synton MDP, Switzerland) in a displacement-controlled mode with a displacement rate of 10 nm/s. The post-testing fracture surface analysis was performed using the same SEM.

Considering the macroscopic brittle behavior of the iron oxide phases reported in the literature [18], and also evident from the fracture tests in this study the fracture toughness measurements were done using linear elastic fracture mechanics (LEFM) formulation. For rectangular cross-sectioned microcantilevers, $\boldsymbol{K_{IC}}$ was measured using the following Eq. 3:

$$K_{IC} = \frac{P_{max}L}{BW^{3/2}} f\left(\frac{a}{B}\right) \quad (3)$$

where $\boldsymbol{P_{max}}$ is the critical load of fracture, $\boldsymbol{L}$ is the distance between the notch and point of loading using spherical indenter. $\boldsymbol{W}$ is the thickness of the sample and $\boldsymbol{B}$ is the width of the sample as annotated in Fig. 1a. The dimensionless geometric factor $\boldsymbol{f\left(\frac{a}{W}\right)}$ was taken from Ref [43,44]. For single crystal magnetite pentagonal microcantilevers, the $\boldsymbol{K_{IC}}$ was measured using the following Eq. 4:

$$K_{IC} = \frac{P_{max}L}{I} \sqrt{\pi a} f\left(\frac{a}{2c}\right) \quad (4)$$

where $\boldsymbol{P_{max}}$, $\boldsymbol{L}$ and $\boldsymbol{a}$ have the same meaning as mentioned in Eq. 3, $\boldsymbol{I}$ is the moment of inertia of the beam and $\boldsymbol{f\left(\frac{a}{2c}\right)}$ is the dimensionless geometric factor calculated from Ref [43,44] and depends on dimensions $\boldsymbol{h}$ and $\boldsymbol{b}$ as shown in Fig. 1b.

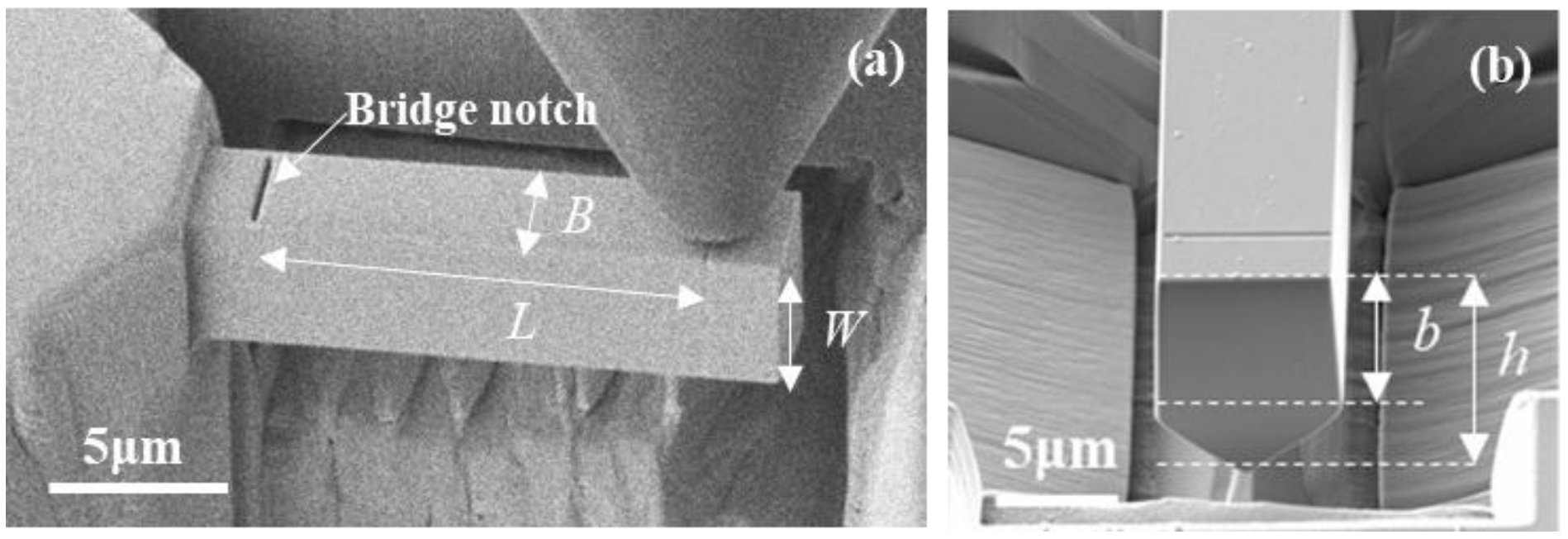

**Fig. 1:** (a) Representative SEM image of a rectangular-shaped microcantilever loaded for *in situ* fracture testing and also highlighting the notations for dimensions and the bridge notch (indicated with an arrow). (b) SEM image showing the cross-section of a pentagonal microcantilever.

## 3. Results

### *3.1. Nanoindentation*

#### *3.1.1. Mechanical properties: Hardness and modulus*

Fig. 2a shows the representative nanoindentation load ***(P)***–displacement ***(δ)*** curves obtained using Berkovich indenter on the (0001) plane of hematite, and (001) plane of magnetite and Wüstite. The measured modulus and hardness for each phase have been plotted in Fig. 2b, with hematite having the highest indentation modulus of 281±3 GPa and hardness of 18.5±0.5 GPa. For magnetite the indentation modulus was measured to be 165±3 GPa which agrees with the elastic modulus reported previously [45–48]. It is important to note that in the case of Wüstite the elastic properties vary with stoichiometry [49]. EPMA measurement showed the stoichiometry of Wüstite to be $Fe_{0.91}O$ (refer to Fig. S1 of Supplementary Information) and the indentation modulus was measured to be 145±4 GPa. The hardness also followed a similar trend with hematite showing the highest hardness at 18.5±0.5 GPa, followed by magnetite at 8.7±0.5 GPa and Wüstite at 7.5±0.4 GPa. With the Berkovich indenter, hematite was the only phase that showed a discernable pop-in at 2.4±0.7 mN load as denoted with arrow in Fig. 2a. This corresponds to a maximum shear stress of 17.2±2 GPa calculated using Eq. 5 [50].

$$\tau_{max} = 0.31 \left(\frac{6E_r^2}{\pi^3 R^2} P_o\right)^{1/3} \quad [5]$$

where, $P_o$ is the pop-in load, $E_r$ is the reduced modulus measured through nanoindentation, $R$ is the radius of the indenter. As no visible crack was observed around the indent on the $(0001)$ plane of hematite, the pop-in maybe associated with possible incipient plasticity events such as dislocation nucleation, although no surface slip trace was visible around the indent. Berkovich indentation was also performed on the $(0\bar{1}10)$ and $(1\bar{1}02)$ planes of hematite which showed cracking during loading preventing reliable determination of $\boldsymbol{H}$ and $\boldsymbol{E}$ for these orientations.

### *3.1.2. Surface slip trace analysis*

The uniform stress field of the spherical indenter makes it ideal to understand the plasticity and cracking behavior of the iron oxides under ambient conditions. The post-deformation SEM micrographs of the spherical indents are shown in Fig. 2c-g, highlighting the presence of different slip traces (denoted by colored dashed lines), twined region (indicated by solid line), pile up and cracking (indicated by arrows) for the 3 iron oxide phases. Of the three orientations indented in hematite, the $(0001)$ orientation revealed no visible surface deformation features as shown in Fig. 2c.

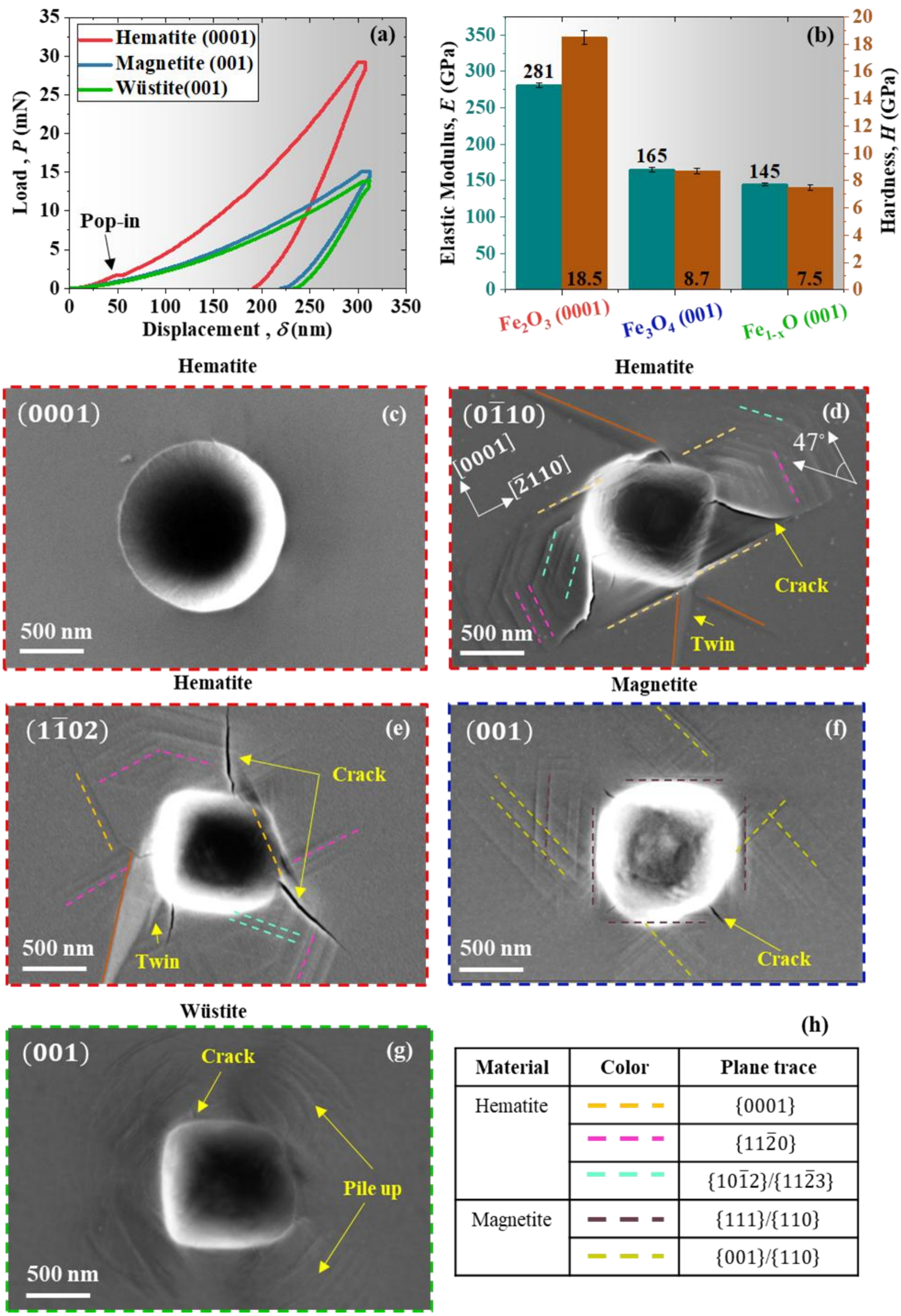

| Material | Color | Plane trace |
|---|---|---|
| Hematite | | {0001} |
| | | {11$\bar{2}$0} |
| | | {10$\bar{1}$2}/{11$\bar{2}$3} |
| Magnetite | | {111}/{110} |
| | | {001}/{110} |

**Fig. 2**: (a) Representative nanoindentation ***P* - *δ*** curves using Berkovich indenter for hematite on (0001) surface, magnetite on (001) surface and Wüstite also on (001) surface. (b) Average ***E*** and ***H***

values obtained from nanoindentation for the three iron oxides; SEM images of spherical indent impressions on (c-e) $(0001)$, $(0\bar{1}10)$ and $(1\bar{1}02)$, surfaces of hematite respectively, and (f,g) $(001)$ surface of magnetite and Wüstite, respectively. (h) color codes indicating crystallographic orientation of the surface traces of the indexed planes in hematite and magnetite. The dashed colored lines represent the slip traces while the solid lines represent the trace of the twin (as further discussed in Sec. 3.1.4).

On the contrary, the $(0\bar{1}10)$ and $(1\bar{1}02)$ orientations of hematite showed multiple features such as slip steps, twin and deformation-induced cracking as annotated in Figs. 2d,e. Such deformation features depend upon the geometry of the indenter, arrangement of surface traces of twinning and slip planes, and symmetry of indentation-induced pile-up. The orientations of the slip traces on the surface were analyzed to identify the slip planes using the crystallographic information from EBSD in conjunction with a Matlab® based slip trace analysis code [38]. A tolerance angle of 3° was maintained between the measured angle of the slip trace with the horizontal and the view plane. For the $(0\bar{1}10)$ plane, the slip trace around the indent can be divided into 3 categories as shown in Fig. 2d. The magenta and yellow lines indicate traces of the $\{11\bar{2}0\}$and $\{0001\}$ planes oriented along $\langle 0001 \rangle$ and $\langle \bar{2}110 \rangle$ direction, respectively. The third set of slip traces, shown in cyan in Figs. 2d,e corresponds to the trace of either $\langle 10\bar{1}2 \rangle$ or $\langle 11\bar{2}3 \rangle$, and could not be unambiguously identified as the surface trace of both these type of planes are aligned along the <c+a> direction which makes 47° to the [0001] direction as shown in Fig. 2d. It is important to note that while calculating the orientation of $\langle c + a \rangle$ one must consider the morphological unit cell which is half of the structural unit cell ($c$=6.885). On the $(1\bar{1}02)$ plane, traces of $\{0001\}$, $\{0\bar{1}10\}$and $\{10\bar{1}2\}$ slip planes were observed around the indent as shown in Fig. 2e. On the $(1\bar{1}02)$ plane the trace for the pyramidal slip confirms to $\{10\bar{1}2\}$ plane without any ambiguity. The slip lines surrounding the spherical indent on the $(001)$ surface of magnetite, shown in Fig. 2e, can be classified into two distinct types. The first set corresponds to the $\{111\}$ and $\{110\}$ slip planes. And the second trace can correspond to both $\{110\}$ and $\{001\}$ planes, both of which have been reported as active slip systems in magnetite [22,51]. In the case of Wüstite, no slip lines were observed; instead, the region surrounding the spherical indent exhibited a pile-up–like feature as shown in Fig. 2g.

### *3.1.3. Cracking and fracture during nanoindentation*

In hematite, no cracking was observed when the $(0001)$ basal plane was indented up to 300 nm. In contrast, cracking occurred on the $(0\bar{1}10)$ and $(1\bar{1}02)$ planes as shown in Fig. 2d,e. In these orientations, cracks formed along the boundary separating regions that experienced plastic deformation from those that remained undeformed or weakly deformed. The regions where slip traces where observed on the surface were considered plastically deformed and the regions without slip traces where considered less or undeformed. On the $(0\bar{1}10)$ plane, the cracks initiated parallel to the $\{0001\}$ or $\{10\bar{1}2\}$ plane close to the circumference of the indent where the tensile stresses are maximum. However, subsequent crack propagation did not follow a single crystallographic plane. For the $(1\bar{1}02)$ plane, crack initiation showed no clear crystallographic alignment, although segments of the crack path were intermittently parallel to the $(0001)$ plane.

For magnetite and Wüstite higher depths of indentation were needed to promote substantial cracking compared to hematite. Fig. 3 a,b shows the indent impression on $(001)$ plane of magnetite and Wüstite after indenting to depths of 1500 nm. For magnetite, radial cleavage cracks originating at the edges of the indent impression (indicated by white arrows) were observed parallel to the $\langle 001 \rangle$ direction, as shown in Fig. 3a (and also in Fig. 2f). A second type of crack was also observed on the $(001)$ plane of magnetite, with the crack developing within the pile up ridges (indicated by yellow arrow). This crack type was more pronounced in Wüstite, as shown in Fig. 3b. In both cases the plastic deformation around the indent impression formed a four-fold symmetry and one crack was associated with each pile-up ridge. These cracks were relatively straight but did not follow a single crystallographic plane.

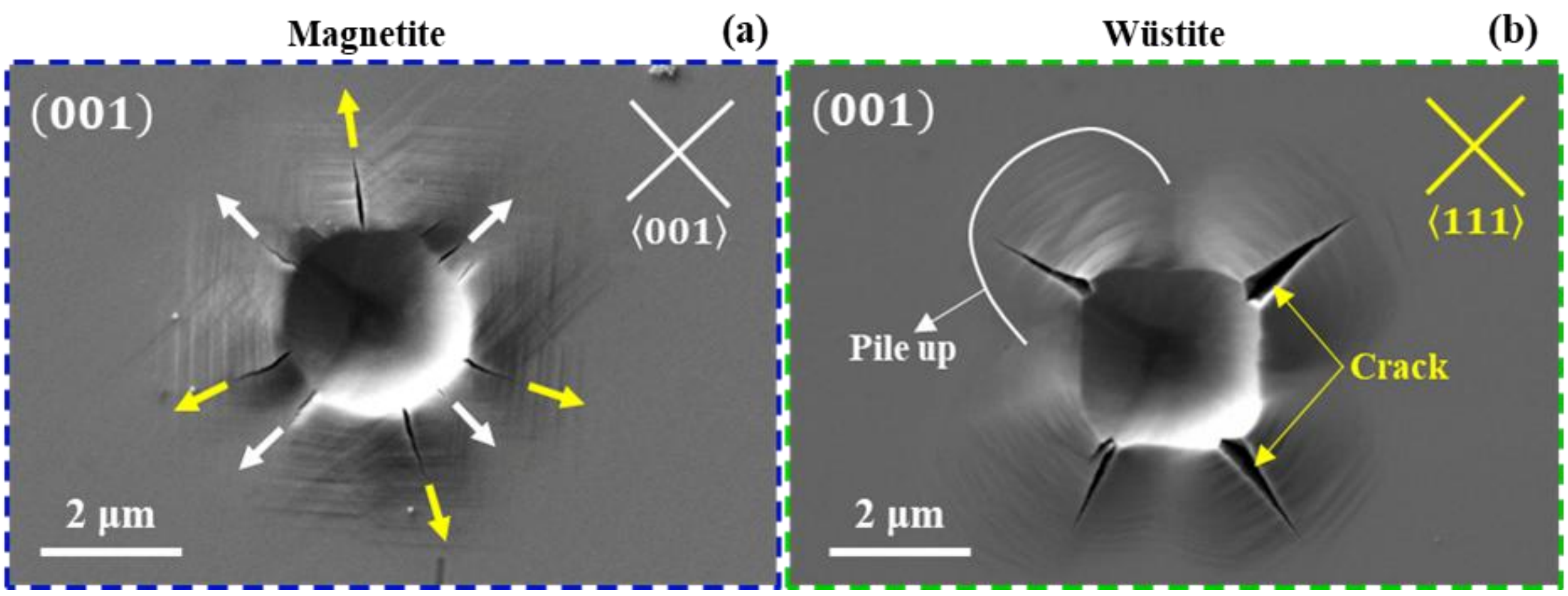

Fig. 3: SEM image of spherical indent impression on $(001)$ plane of (a) magnetite and (b) Wüstite, upto depths of 1500 nm. Surface directions are included.

### *3.1.4. Twinning in hematite*

Twinning was observed only in hematite on the $(0\bar{1}10)$ and $(1\bar{1}02)$ plane as shown in Fig. 2 d,e. The $(0001)$ plane, showed no evidence of twinning around the indent as shown in Fig. 2c. Similar observations have also been reported for the crystallographically similar material sapphire, for which no deformation features were reported on the $(0001)$ surface [52]. Needle/lenticular-shaped twin was observed around the indent impression on the $(0\bar{1}10)$ and $(1\bar{1}02)$ surfaces (denoted by solid line in Fig. 2d,e). The trace of the twin aligned with the orientation of *R* twin as determined by Tymiak and Gerberich on both $(0\bar{1}10)$ and $(1\bar{1}02)$ planes [52]. The crystallography of the *R* twin was determined using an angle-axis pair which relates the parent with the twinned crystal keeping one of the planes undistorted, also known as K1 plane [53]. EBSD analysis of the indent on $(0\bar{1}10)$ plane is shown in Fig. 4a, where the misorientation angle between the parent and the twinned crystal was measured to be 85±1$^o$ (plotted in Fig. 4b). Once the misorientation angle is known the axis of rotation can be determined from Fig. 4c, which shows an inverse pole figure (IPF) for rotation axes for parent hematite in a crystal coordinate system. The IPF is taken only for the interval of rotation angle from 83$^o$ to 87$^o$. It shows a strong concentration around direction $\langle\bar{2}021\rangle$ which is the rotation axis in this case. Bursill *et al.* [54] have studied the *R* twin in hematite and have reported that rotation of 84.8$^o$ along $\langle 02\bar{2}1\rangle$ direction can bring the parent and twin into coincidence. This is consistent with the EBSD results shown in Fig. 4. As the invariant plane K1, is the undistorted plane shared between parent and twinned crystal the intensity spots of parent and twinned crystal in the pole corresponding to the K1 plane would coincide with each other. This was observed in the $\{01\bar{1}2\}$ poles of the parent and twinned crystal (Fig. 4d). Hence the twinned crystal is rotated by 85$^o$±1$^o$ along the $\langle\bar{2}021\rangle$ axis keeping the $\{01\bar{1}2\}$ plane undistorted with respect to the parent crystal. This angle-axis pair and misorientation angle was used to identify the rhombohedral twin boundary in the TSL OIM software which traced the twin/parent interface as shown in Fig. 4e.

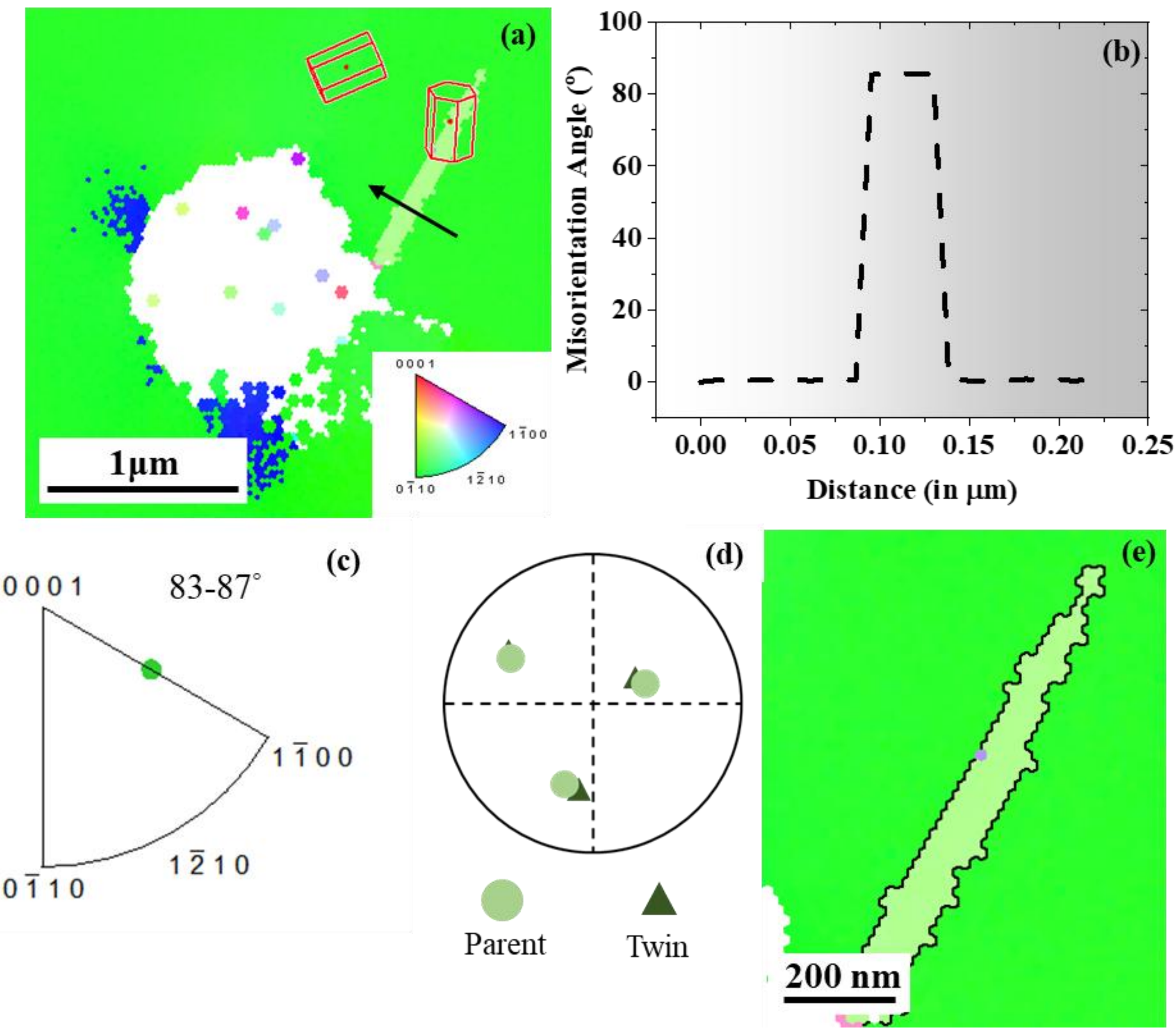

**Fig. 4:** (a) Electron backscattered diffraction (EBSD) analysis of the deformation twin around the spherical indent impression on (0$\bar{1}$10) plane of hematite. (b) Point to origin misorientation profile taken along the twin cross section as shown in (a). (c) Inverse pole figure (IPF) corresponding to EBSD in (a) showing the rotation axis in crystal coordinate system. The IPF is taken for the interval of rotation angle of 83°-87°. (d) Pole figure corresponding to {01$\bar{1}$2} showing superposition of intensity spots from the parent and twinned crystal. (e) Highlighted twin boundary in the TSL OIM software using determined angle-axis pair.

### *3.2. Microcantilever fracture toughness testing*

#### *3.2.1. Fracture of low index planes*

In brittle materials, low-index planes typically represent regions of weaker atomic bonding and are therefore more susceptible to cleavage failure [39,55,56]. Based on available literature for iron oxides and other ceramics with similar crystal structures, the {0001} plane of hematite and the {001} plane of magnetite and Wüstite were selected as the notch planes [57], to elucidate the fracture behavior. Fig. 5a presents the $\boldsymbol{P}$ vs. $\boldsymbol{\delta}$ curves for all three iron oxide microcantilevers, with notches along {0001} and {001} planes. The EBSD showing the orientation of notch with respect to the crystal structure for the three iron oxide phases is presented in Fig. S4 (Supplementary Information). The geometric details of the low index orientation microcantilevers are provided in Table S1 and S2 (Supplementary Information). Each curve exhibits a linear elastic behavior, followed by a sudden load drop at the peak load ($\boldsymbol{P_{max}}$) (denoted by the asterisk * symbol in Fig. 5a), indicative of unstable crack propagation and brittle fracture. The fracture toughness ($\boldsymbol{K_{IC}}$) was calculated using Eq. 3 for the rectangular beams of hematite and Wüstite, while Eq. 4 was applied for the pentagonal shaped cantilevers of magnetite. The measured fracture toughness values, reported in Fig. 5b, indicate that hematite has the highest fracture toughness at 3.6±0.2 MPa$\sqrt{m}$, followed by magnetite at 2.6±0.1 MPa$\sqrt{m}$ and Wüstite at 2.0±0.2 MPa$\sqrt{m}$. Crack propagation in all three cases was recorded through *in-situ* video monitoring. Multiple instances of crack deviation were observed in hematite, leading to a tortuous crack path, as shown in Fig. 5c. In contrast, magnetite and Wüstite exhibited microscopically straight cracks confined to the plane of the notch, i.e. {001} plane, without any noticeable crack deviation as shown in Fig. 5d and e.

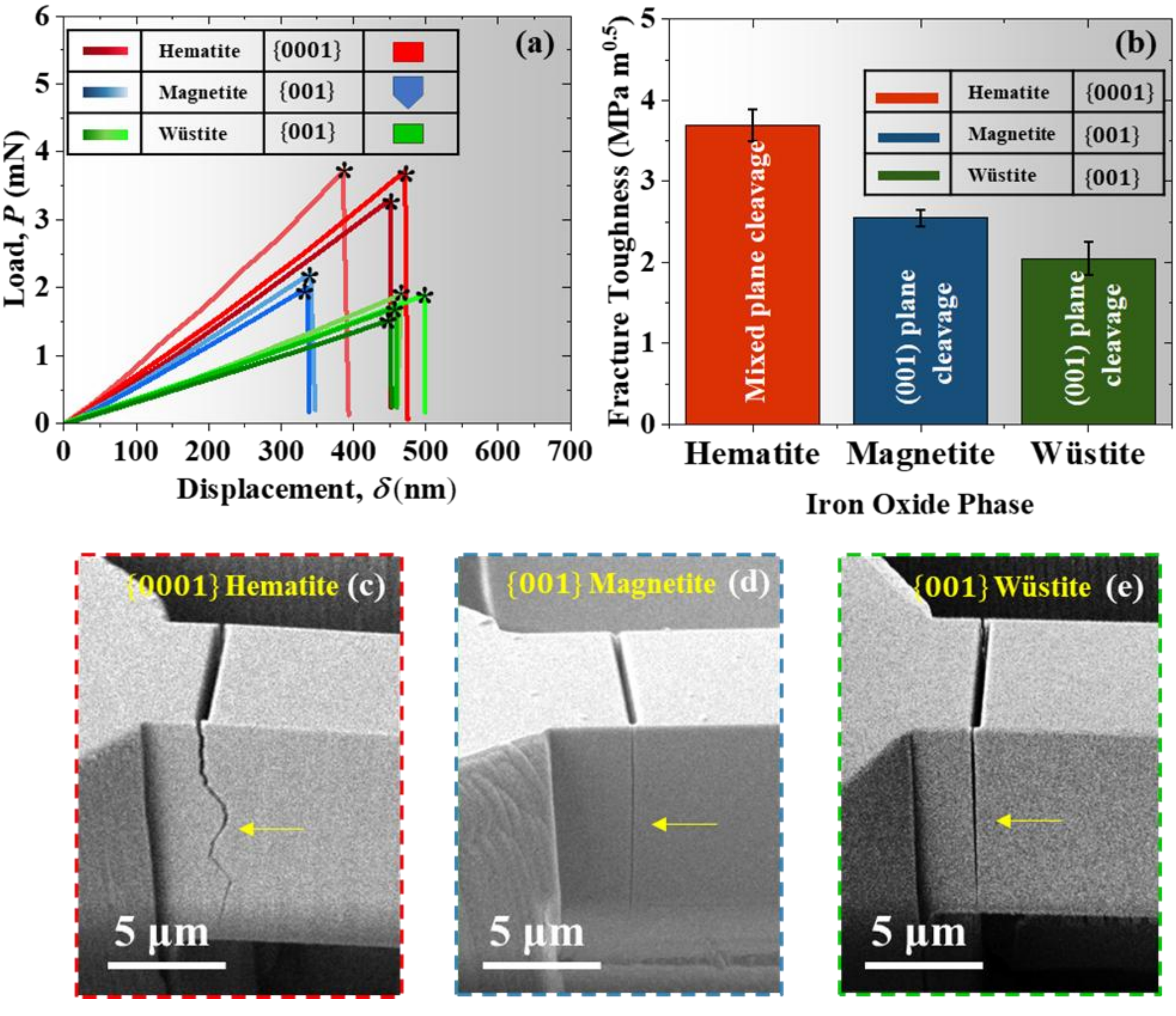

**Fig. 5:** (a) Load-displacement ***(P - δ)*** curves for fracture toughness measurement of low index planes of the 3 iron oxide phases using microcantilever bending. (b) Fracture toughness calculated from the critical crack initiation load denoted by (*) from (a) using Eq. 3 for hematite, Wüstite and Eq. 4 for magnetite. (c-e) Post fracture SEM images showing crack propagation during fracture in $\{0001\}$ plane of hematite, $\{001\}$ plane of magnetite and $\{001\}$ plane of Wüstite, respectively.

### *3.2.2. Fracture of high index planes*

In ionic and partially ionic materials high index planes are generally polar in nature resulting in more complex bonding arrangements, often characterized by unsaturated or distorted surface bonds compared to low index planes which are usually non-polar in nature. This structural complexity raises the surface energy of the high index planes [58]. Without large-scale plastic deformation, the increased surface energy enhances resistance to cracking, thus contributing to their higher fracture toughness. Therefore, high-index planes can be considered as suitable orientations for evaluating the upper limit of fracture toughness for iron oxide phases. The specific notch-plane was aligned with different high index orientations of hematite, magnetite, and Wüstite as listed in Table 2, measured from EBSD (refer to Fig. S5 of Supplementary Information for details). Geometric details for high index oriented microcantilevers are provided in Tables S3 and S4. Fig. 6(a) shows the ***P–δ*** response for the three iron oxide phases with the notch plane aligned along the high index planes. The fracture toughness values are summarized in Fig. 6(b). For hematite, the measured fracture toughness of the high-index plane was 3.4±0.4 MPa$\sqrt{m}$ similar to that of the low-index (0001) plane. Conversely, magnetite and Wüstite showed fracture toughness of 4.4±0.3 MPa$\sqrt{m}$ and 2.8±0.1 MPa$\sqrt{m}$ respectively for the high-index orientations which was higher compared to their low-index counterparts. Specifically, the fracture toughness of magnetite increased by around 80%, while that of Wüstite increased by about 40%. Crack propagation paths were observed through *in-situ* video recordings, with representative snapshots immediately after crack initiation at the peak load shown in Fig. 6c–e. In hematite and magnetite, cracks deviated notably from the notch plane, following non-planar paths (Fig. 6c,d). In contrast, in Wüstite, the crack moved along a relatively straight path, closely aligned with the notch plane (Fig. 6e). Further discussion of the fracture surfaces of the three iron oxide phases is provided in Section 4.3.

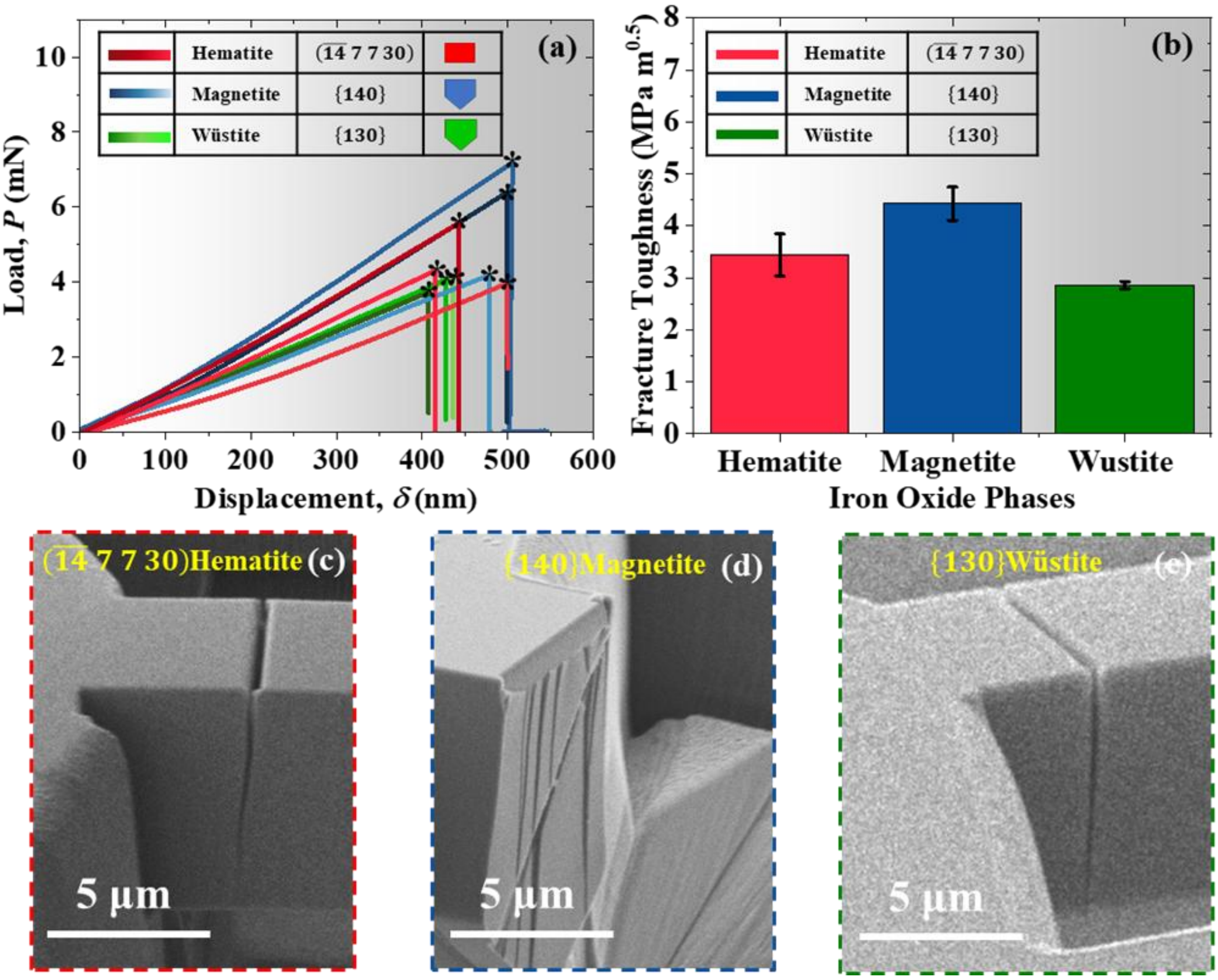

**Fig. 6:** (a) Load-displacement ***(P-δ)*** curves for fracture toughness measurement along high index orientation of the three iron oxide phases using microcantilever bending; the asterisks (*) denote critical load for fracture; (b) Fracture toughness values measured from the critical crack initiation load from (a) using equations 3 & 4. (c-e) Post fracture SEM images of crack propagation during fracture of hematite, magnetite and Wüstite respectively.

### *3.2.3. Fracture along oxide-gangue interface*

The interface fracture between magnetite and (Al, Si, Mg)–O based gangue phase was studied using micro-cantilever fracture testing (Fig. 7a). The crystallographic orientation of single-crystal magnetite relative to the interface is illustrated by EBSD mapping (Fig. 7b). For comparison single-crystal magnetite along the same orientation (120) was also tested. The load-displacement ***(P-δ)*** behavior for the interface fracture testing and the (120) oriented magnetite is compared in Fig. 7c. The fracture toughness of the interface was evaluated on two specimens, yielding values of 0.6 and 0.8 MPa$\sqrt{m}$. These values are nearly an order of magnitude lower than the fracture toughness of the (120) plane of magnetite, which was determined to be 3.9 ± 0.3 MPa$\sqrt{m}$. For the magnetite-gangue composite sample, the crack propagation happened along the interface between the gangue particle and magnetite as shown in Fig. 7d. The corresponding *in-situ* fracture video is provided in Supplementary Video 1.

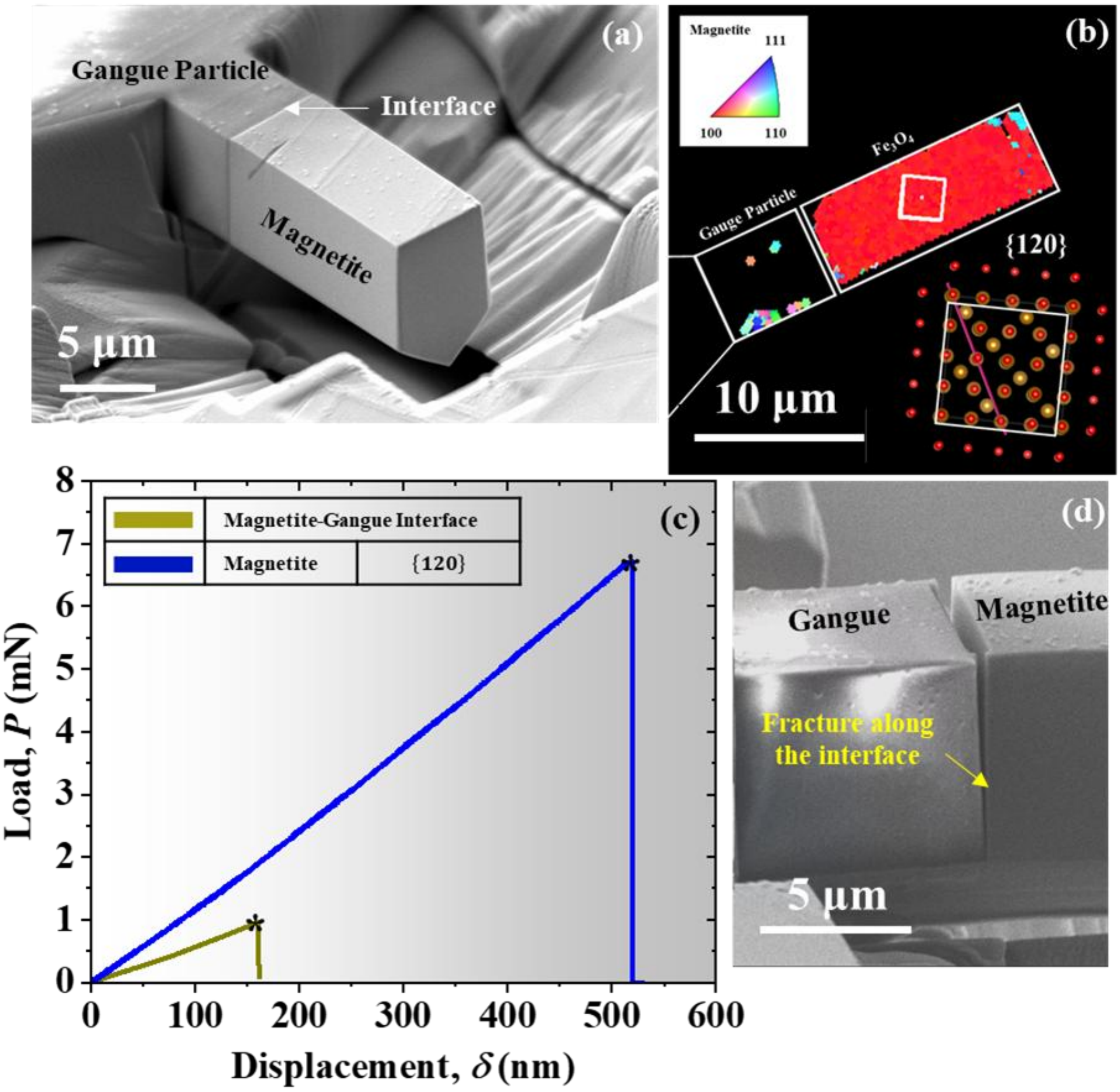

**Fig. 7:** (a) SEM image of microcantilever with interface between a gauge particle and single crystal magnetite. (b) EBSD map of the microcantilever showing the orientation of magnetite. (c) Load - displacement ***(P vs δ)*** behavior from fracture toughness measurements along the interface between the ganguee particle and magnetite, and single crystal magnetite of same orientation. (d) SEM image showing final fracture along the interface between the gangue particle and magnetite.

## 4. Discussion

### *4.1. Comparison of nano-mechanical properties, indentation cracking and plasticity in iron oxide phases*

#### *4.1.1. Elastic Modulus*

The elastic properties of the three iron oxide phases have been extensively studied in literature due to their crucial role in mitigating spalling during the hot rolling of steel slabs [45]. The elastic modulus values exhibit a broad range for both hematite and magnetite, spanning from 200 to 270 GPa for hematite [45,48,59–63] , while those for magnetite range from 160 to 210 GPa [46,48,59,60,64–67]. This variability arises from the differences in sample types tested—ranging from thin oxide films to natural bulk ores—as well as the use of diverse measurement techniques, each contributing to the scatter in the reported values. The majority of nanoindentation-based measurements are performed on oxide scales, which commonly possess a gradient in microstructure—comprising of an outer hematite layer transitioning into magnetite and subsequently Wüstite [45,59]. These heterogeneous microstructures often contain defects introduced by volumetric changes during phase transformations, which can negatively influence the modulus values. Chicot *et al.*, [48] measured the elastic modulus of a naturally grown polycrystalline hematite to be 237±4 GPa using nanoindentation, which compares well with the elastic modulus of 281±3 GPa reported in this study. When compared to hematite, magnetite showed 40% lower indentation modulus which was measured to be 165±3 GPa, well within the range of previously reported literature values [45–48,66,67]. Crystallographic data show that the $Fe^{3+}$–O bond lengths in α-$Fe_2O_3$ are in the range ~1.97–2.12 Å in octahedral coordination, whereas in $Fe_3O_4$ the octahedral $Fe^{2+}$–O distances are larger (2.02-2.16 Å) due to the larger ionic radius of $Fe^{2+}$ compared with $Fe^{3+}$. The shorter, stronger $Fe^{3+}$–O bonds in hematite may contribute to its higher elastic stiffness compared to magnetite [68].

The elastic modulus of Wüstite is known to vary as a function of cation vacancy concentration. For $x$>0.98, the elastic modulus increases as cation vacancies decrease, with stoichiometric FeO exhibiting theoretically predicted bulk modulus of 185 GPa [49]. In contrast, for cation concentrations below 0.98, the elastic modulus remains scattered between 130-160 GPa, [49,69,70]. In this study, the cation ($Fe^{2+}$) concentration was determined to be 0.91 using EPMA and the elastic modulus was measured to be 145±3 GPa. Magnetite, as discussed previously, contains both $Fe^{3+}$–O and $Fe^{2+}$–O bonds, with the latter characterized by longer bond lengths and consequently lower bond stiffness. Wüstite, in contrast, consists exclusively of $Fe^{2+}$–O bonds and, when coupled with its high concentration of cation vacancies ($Fe_{0.91}O$), exhibits a markedly reduced elastic stiffness.

#### *4.1.2. Hardness and Slip Activation*

Hardness is a measure of the material's resistance to plastic deformation and in low-symmetry crystals such as hematite, it is often controlled by the most difficult deformation mechanism to activate. According to literature [17], basal and prismatic slip constitute the primary room temperature deformation modes in hematite, accompanied by rhombohedral and basal twinning. In the present study, slip traces corresponding to the $\{0001\}$ and $\{11\bar{2}0\}$ planes were observed on the $(0\bar{1}10)$ and $(1\bar{1}02)$ surfaces (Fig. 2c,d), confirming the activation of basal and prismatic slip. In addition, both surfaces exhibited $\{10\bar{1}2\}$ slip traces associated with *R*-type pyramidal slip, a mechanism which has been previously reported in α-alumina, having similar crystal structure as hematite. The relatively high hardness of hematite (~18.5±0.5 GPa) may be attributed to the activation of pyramidal slip, which requires higher stresses than the primary deformation modes, particularly at room temperature[71–73].

Magnetite has a cubic spinel structure which deforms similar to an FCC metal through dislocation glide on $\{111\}\langle 1\bar{1}0\rangle$ as primary slip system at room temperature. Further, Charpentier [74], in studies on plasticity of synthetic magnetite, identified $\{100\}\langle 0\bar{1}1\rangle$ as secondary slip system active at room temperature. In the current study slip traces corresponding to both $\{111\}$ and $\{100\}$ were observed as shown in Fig. 2f, although they cannot be unambiguously separated from the traces of $\{110\}$ planes. From magnetite to Wüstite cubic crystal structure is retained, but a shift in cation coordination in the vacancies changes the structure from inverse spinel to rocksalt which deforms primarily by $\{110\}\langle 1\bar{1}0\rangle$ slip [18,20]. In the case of Wüstite, the region surrounding the spherical indent on the $\{100\}$ surface exhibited 'pile-up' like feature (Fig. 2f). Unlike the well-defined slip traces observed in hematite and magnetite, the pile-up features in Wüstite were curved and lacked a clear crystallographic alignment, impeding the extraction of slip-system information from these surface features. This was also observed in indents made at lower depths. This contrast reflects fundamental differences in the accommodation of plastic deformation: materials in which plastic deformation is highly localized and accommodated by a limited number of slip systems tend to exhibit discrete slip traces, whereas materials that deform more homogeneously through the activation of multiple slip systems, results in pile-up formation around the indent [38,75].

The Peierls–Nabarro stress for the $\{111\}\langle 1\bar{1}0\rangle$ slip system in magnetite was estimated as approximately G/5, whereas for $\{110\}\langle 1\bar{1}0\rangle$ slip in Wüstite it was calculated to be G/15 using Eq. 6, explaining the lower hardness in Wüstite compared to magnetite. Additionally, along the reduction sequence from hematite to Wüstite, the Fe:O ratio increases from 0.66 to 0.75 and further to 0.91 (for this study), indicating a progressive shift toward weaker covalent bonding. This trend is supported by band gap values, which decrease from 2.6 eV in hematite to 1 eV in Wüstite, implying an increase in electron density or metallic bonding, which also contributes to the decrease in hardness in Wüstite relative to magnetite [76–78].

$$\tau_{PN} = \frac{2G}{(1-\upsilon)} exp\left(\frac{-2\pi d}{b}\right) \quad [6]$$

Where $\tau_{PN}$ is the Peirels-Nabarro stress (critical shear stress for dislocation motion), $G$ is shear modulus, $\upsilon$ is Poisson's ratio, $d$ is interplanar spacing of the slip plane and $b$ is the magnitude of Burgers vector.

### *4.1.3. Cracking during spherical nanoindentation*

All three iron oxide phases exhibited distinct cracking behavior under spherical indentation, reflecting differences in their fracture behavior and deformation mechanisms. In hematite, the cracks observed on the $(0\bar{1}10)$ and $(1\bar{1}02)$ plane delineate the boundary between the plastically deformed region and the surrounding region that experienced little or no plastic deformation. In case of hematite, which possesses a corundum (R-3c) structure with pronounced crystallographic anisotropy, the distribution of resolved shear stresses is highly heterogeneous. Consequently, plastic deformation is preferentially activated in specific regions around the indent while neighboring regions remain largely elastic. Similar spatially heterogeneous deformation behavior has been predicted by Tymaik for α-alumina [52], which is isostructural with hematite, where the arrangement of slip traces and twins under indentation reflects the anisotropic distribution of resolved shear stresses. The crack nucleation has occurred along crystallographically favorable planes such as $\{0001\}$ or $\{10\bar{1}2\}$, as observed in Fig 2d,e. However, localized plastic deformation generates strain gradients at the boundary of the plastic zone, leading to incompatibility and hence further crack propagation happens along the boundary separating the plastically deformed and undeformed regions.

In magnetite, two distinct cracking modes were observed. The first corresponds to radial cracks emanating from the periphery of the indent impression and propagating along low-energy cleavage planes. The traces of these cracks were aligned with the $\langle 001 \rangle$ direction, both of which are known cleavage planes in magnetite with surface energies ($\gamma_{sf}$) of approximately 1.5 and 1.8 $J/m^2$ [65], respectively. A second cracking mode, referred to here as "pile-up–induced cracking", was observed in magnetite and occurred more prominently in Wüstite. In this case, cracks formed within the pile-up ridges surrounding the indentation impression. Both magnetite and Wüstite exhibit higher dislocation mobility compared to hematite, enabling more plastic deformation during nanoindentation. This enhanced plasticity promotes outward material flow and the formation of pronounced pile-up around the indent. The upward displacement of material within the pile-up ridge generates tensile and bending stresses along the ridge crest during continued deformation which can, result in the observed cracks that propagate through the pile-up region and appear as tearing of the ridge.

## *4.2. Comparison of fracture behavior of the iron oxides*

As discussed in Results Section the fracture toughness testing of the low index and random high index planes of the three iron oxide phases exhibited linear elastic fracture and the measured $K_{IC}$ values have been summarized in Fig. 5b and 6b. For hematite, the measured fracture toughness ranged between 3.4 to 3.6 MPa$\sqrt{m}$, which is in reasonable agreement with the reported value of 2.3 ± 1 MPa$\sqrt{m}$ in the literature for commercial iron ore pellets [8]. The comparatively lower fracture toughness of commercial pellets can be attributed to the presence of porosity and gangue particles, which act as stress concentrators and facilitate crack initiation and propagation, unlike the defect-free single crystal microcantilever examined in the present study.

The low-index fracture planes of magnetite and Wüstite exhibited clean cleavage failure with minimal crack deviation as evident from the fracture surfaces shown in Fig. 8c and d. This is further supported by the straight crack path along the notch plane observed from the snapshot taken from the *in-situ* video captured during the tests (refer Fig. 5d,e). However, the same was not true for hematite which exhibited crack deviation leading to a tortuous crack propagation as evident from Fig. 8a, b. The fracture surface of hematite shows multiple facets being formed during the cracking process. Fig. 8a shows the side view of the cantilever revealing the crack path during fracture. Further crystallographic analysis of the facets on the fracture surface using Vesta® crystallographic visualizer, as shown in inset in Fig. 8b, highlight that the crack path follows the rhombohedral plane $\{10\bar{1}2\}$ with the formation of multiple rhombohedral facets on the crack plane. The trace of the rhombohedral planes on the view plane i.e. $(11\bar{2}0)$ in Fig. 8a and $(0001)$ for Fig. 8b is highlighted in yellow. This deviation of the crack path from $(0001)$ to $\{10\bar{1}2\}$ plane is due to the low surface energy configuration of the $\{10\bar{1}2\}$ plane. Robertson and Manning have reported the surface energy for $\{10\bar{1}2\}$ and $\{0001\}$ plane to be 1.95 J/m$^2$ and 2.25 J/m$^2$ respectively [65]. Hence, during the crack propagation there was a preference for the crack to deviate to the low energy $\{10\bar{1}2\}$ plane. Such a tendency to crack along the $\{10\bar{1}2\}$ rhombohedral plane has also been reported in α-alumina which shared the same crystal structure as hematite [79]. This tortuous crack propagation can lead to toughening. Hence the hematite $(0001)$ notch plane fracture toughness was higher compared to the other two iron oxide phases.

In the case of magnetite and Wüstite which showed clean crack propagation along the cubic plane, the fracture toughness scaled with the corresponding surface energy values. However, the experimentally measured critical fracture energies for magnetite and Wüstite were found to be approximately 10–14 times higher than the theoretical values estimated from the corresponding surface energy ($\gamma_{sf}$) values. To evaluate whether this discrepancy could arise from small-scale plasticity at the crack tip, unnotched microcantilever bending tests were performed to determine the fracture strength and estimate the critical flaw size (Section 9 of the Supplementary Information). The calculated critical flaw dimension was comparable to the notch depth, indicating negligible or no plastic deformation prior to fracture. These unnotched fracture tests were conducted on high-index oriented microcantilevers of hematite and Wüstite. The absence of detectable small-scale plasticity suggests that the elevated experimentally measured fracture energy does not originate from plastic dissipation near the crack tip. Instead, the higher fracture energy may arise from other mechanisms such as lattice trapping effects [80–82], kinetic barriers associated with crack propagation [81,82], or local inhomogeneities and geometric irregularities at the notch tip. While the exact reasoning for this cannot be validated in this study, the fracture surface corresponding to {001} plane of magnetite and Wüstite shows ratchet marks at the crack front near the notch tip (Fig. 8c and d). These ratchet marks are step-like features observed on the fracture surface that are formed when the microcracks initiate at slightly different locations along the notch front and then coalesce during crack propagation [83]. Their presence near the notch tip indicates that after crack nucleation at the tensile bridges the propagation was not uniform at the notch front with multiple competing sites. One plausible suspicion for this non-uniformity could be redeposition of sputtered material during the notch milling. Once the crack propagates, it aligns itself to the (001) plane, leaving a ratchet-free clean planar cleavage surface at the bottom half of the fracture surface.

For the high-index orientations, the fracture surfaces of hematite, magnetite, and Wüstite are presented in Fig. 8e-g. In hematite and magnetite, the crack paths deviated from the notch plane during fracture (Fig. 6c,d), whereas in Wüstite the crack propagated predominantly along the notch plane (Fig. 6e). In all three phases, pronounced ratchet markings were observed at the crack tip (Fig. 8c-g). In magnetite, these ratchets extended across the entire fracture surface, suggesting a reduced driving force for single-planar cleavage compared to the low-index orientation. The enhanced toughness in hematite and magnetite, therefore, arises from a combination of crack deviation and ratcheting. In contrast, Wüstite exhibited intense ratcheting localized near the notch; however, the crack path ultimately coalesced along the (130) plane, similar to its behavior in low-index orientations. This suggests that the enhanced toughness observed in high-index (130) Wüstite arises from a combination of the intrinsically higher surface energy associated with high-index planes and the additional energy dissipation introduced by ratcheting deformation.

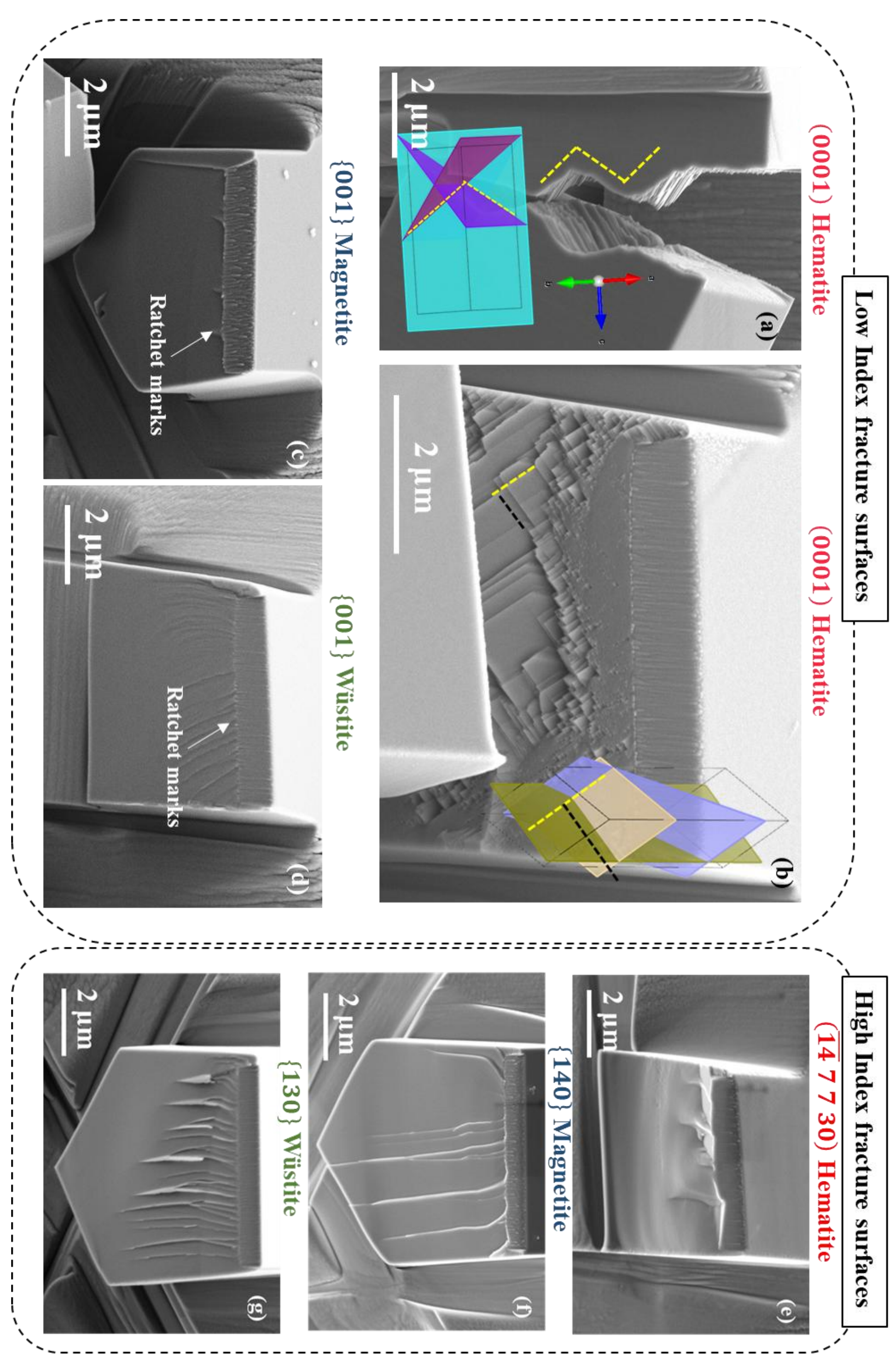

**Fig. 8:** (a & b) SEM image showing the crack propagation in (0001) oriented hematite. (b) (0001) fracture surface showing facets aligning to rhombohedral planes of hematite. (c & d)_SEM image showing fracture surface of (001) oriented magnetite and Wüstite highlighting the ratchet marks. (e-g)

SEM image showing the fracture surfaces of high index oriented microcantilever of hematite, magnetite and Wüstite, respectively.

*4.3. Implication of the mechanical properties on structural integrity of iron oxide pellet*

The above discussions give a clear picture of the changes in micromechanical deformation and fracture behavior of the iron oxides at different stages of reduction. As discussed in the Introduction, the elevated gas flow rates in the HyDR process intensify mechanical degradation, primarily through abrasion arising from collisions between pellets and between pellets and the reactor walls. The predominant damage mechanisms are surface wear and chipping, which together account for the majority of iron ore fines generated during handling and the descent of feedstock along the reduction column. Ghadiri and Zhang [12] demonstrated that these interactions can be modelled as indentation-induced fracture, linking fines generation to the structural integrity of the pellets through two material-dependent parameters: $\boldsymbol{H}$ and $\boldsymbol{K_c}$. However, since these properties do not exhibit identical scaling relationships with attrition intensity, a more comprehensive parameter is required for comparison. In this study, the brittleness index, $\boldsymbol{Q}$ [14] is employed to evaluate the relative attrition behavior of the three iron oxides. The brittleness index, defined by Eq. 9, has unit of inverse length (μm$^{-1}$), reflecting a characteristic length scale governing the transition between plastic deformation and fracture; materials with higher values exhibit fracture at smaller deformation volumes during indentation loading. In this framework, hardness ***(H)*** represents the deformation energy per unit volume, while $\frac{K_{Ic}^2}{E}$ corresponds to the fracture surface energy per unit area. Thus, ***(Q)*** reflects the balance between deformation and fracture processes.

$$Brittleness\ index\ (\boldsymbol{Q}) = \frac{H\,E}{K_{Ic}^2} \qquad [9]$$

Fig. 9 compares the $\boldsymbol{Q}$ values for the 3 iron oxide phases. The high $\boldsymbol{H}$ (18.5±0.5 GPa) and $\boldsymbol{E}$ (281±3 GPa) results in a high $\boldsymbol{Q}$ (≈380-430 μm$^{-1}$) value for hematite indicating a relatively smaller deformation volume compared to magnetite ($Q$ ≈60-230 μm$^{-1}$) and Wüstite ($Q$ ≈130-240 μm$^{-1}$), making it more susceptible to attrition through chipping. In contrast, the lower $\boldsymbol{H}$ and $\boldsymbol{E}$ of magnetite ($\boldsymbol{H}$=8.7±0.5 GPa, $\boldsymbol{E}$=165±3 GPa) and Wüstite ($\boldsymbol{H}$=7.5±0.4 GPa, $\boldsymbol{E}$=145±4 GPa,) results in relatively lower $\boldsymbol{Q}$ and a greater propensity for plastic accommodation. Nevertheless, their relatively low fracture toughness ultimately allows crack initiation once the accumulated stresses within the pile-up regions exceed the fracture resistance.

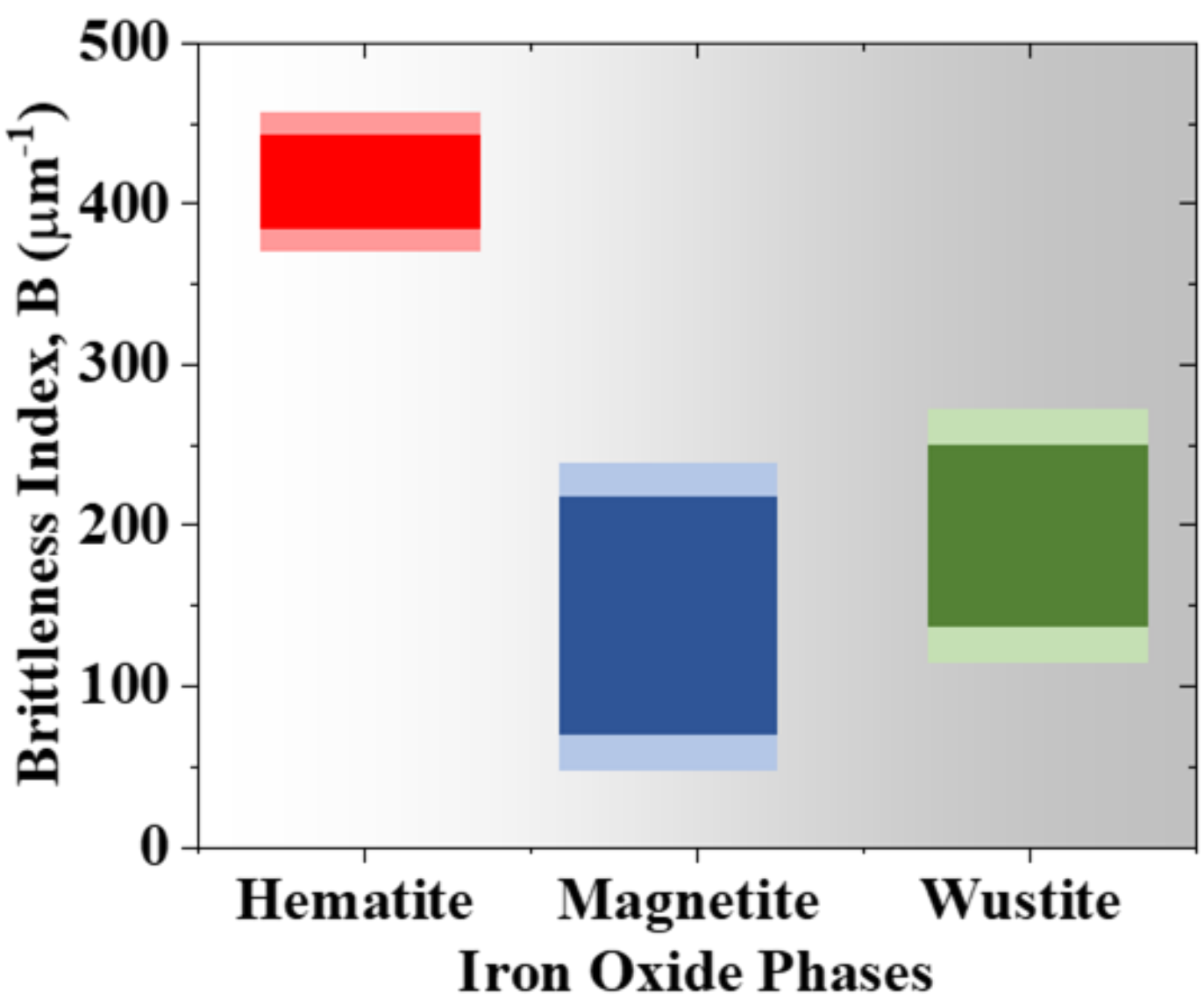

**Fig. 9:** Comparison of the brittleness index (***Q***) for hematite, magnetite, and Wüstite. The lighter shaded regions above and below each symbol represent the uncertainty arising from errors in the measured mechanical properties.

## 5. Conclusion

The fundamental deformation and fracture behavior of the 3 iron oxide phases were investigated at room temperature. The measured mechanical properties—hardness, elastic modulus, and fracture toughness—constitute important input parameters for large-scale material models aimed at predicting fines loss during hydrogen-based direct reduction (HyDR). The principal outcomes of this work are summarized below.

- Hematite exhibits the highest stiffness and hardness ($\boldsymbol{E} \approx 281$ GPa, $\boldsymbol{H} \approx 18.5$ GPa), followed by magnetite ($\boldsymbol{E} \approx 165$ GPa, $\boldsymbol{H} \approx 8.7$ GPa) and Wüstite ($\boldsymbol{E} \approx 145$ GPa, $\boldsymbol{H} \approx 7.5$ GPa), reflecting the progressive transition from strong $Fe^{3+}$–O bonding to $Fe^{2+}$–O bonding and increased cation vacancy defect concentration, in Wüstite, along the reduction sequence.
- Hematite exhibits cracking along the boundary between plastically deformed and less deformed regions, magnetite exhibits mixed radial and pile-up induced cracking, while Wüstite predominantly shows pile-up–induced cracking due to more homogeneous plastic flow.
- Microcantilever fracture testing along low index planes show highest toughness in hematite (~3.5 MPa√m) due to crack deviation and faceting, while magnetite (~2.6 MPa√m) and Wüstite (~2.0 MPa√m) fracture by single-plane cleavage. In high-index orientations of magnetite and Wüstite, enhanced fracture toughness was observed compared to low-index planes. In magnetite, this toughening is primarily associated with crack deviation and ratcheting mechanisms, whereas in Wüstite, where fracture occurs by single-plane cleavage, the increased toughness is attributed to higher surface energy of the cleavage plane combined with ratcheting effects.

- Hematite exhibits the highest brittleness index ($\boldsymbol{Q} \approx$ 380–440 $\mu m^{-1}$), indicating highly localized plastic deformation and crack initiation at the boundary between plastically deformed and undeformed regions, which may promote increased attrition. In contrast, magnetite ($\boldsymbol{Q} \approx$ 60–230 $\mu m^{-1}$) and Wüstite ($\boldsymbol{Q} \approx$ 130–240 $\mu m^{-1}$) show lower $\boldsymbol{Q}$ values due to their lower hardness, enabling greater plastic accommodation and stress redistribution, thereby reducing their susceptibility to attrition.
- In low concentrate iron oxide pellets, gangue-iron oxide interfaces tend to have 3-8 times lower fracture toughness (0.6-0.8 MPa$\sqrt{m}$) compared to bulk single crystals (2-4.5 MPa$\sqrt{m}$ ) as observed from the gangue-magnetite interface results, highlighting weak interfaces as critical sites for chipping and attrition in iron oxide pellets under indentation type loading.

While the present study, conducted on model single crystals, provides valuable insights into the micromechanical behavior of individual iron oxide phases, caution is required when extrapolating these results to industrial pellets, which contain significant microstructural heterogeneities. Furthermore, the experiments were performed under quasi-static loading conditions, whereas pellet–pellet and pellet–wall interactions within reduction reactors occur under much higher speed better mimicked by high strain-rate conditions.

## 6. Acknowledgement

Authors gratefully acknowledge funding from the KSB Stiftung of project FeOx-HOTMECH. S.S. acknowledges Irina Wossack for help with EPMA. S.S. is also thankful to Andreas Sturm for timely help with operation of FEI Helios CX G3 PFIB.

*Supplementary information for*

# Resolving room temperature microscale fracture and plasticity of iron oxide along the cascade of iron ore reduction via nanoindentation and microcantilever bending

Shreehard Sahu*, James P. Best, Gerhard Dehm and Anwesha Kanjilal*

*Max Planck Institute for Sustainable Materials, 40237 Düsseldorf, Germany*

*Corresponding authors: s.sahu@mpi-susmat.de, a.kanjilal@mpi-susmat.de*

**1: EPMA result of Wüstite ($Fe_{1-x}O$)**

The EPMA was conducted on (001) single crystal of Wüstite using a JEOL JXA-iSP 100 EPMA at 15 keV at 12 different spots. The weight percentage of Fe and O is measured which is later converted to atomic percentage for the stoichiometry of $Fe_{1-x}O$.

The average atomic % of Fe and O was measured to be 47.8% and 52.2% respectively which results in a stoichiometry of $Fe_{0.91}O$ or $Fe_{1-0.09}O$.

**2: Energy dispersive spectroscopy (EDS) of gangue particle-magnetite interface**

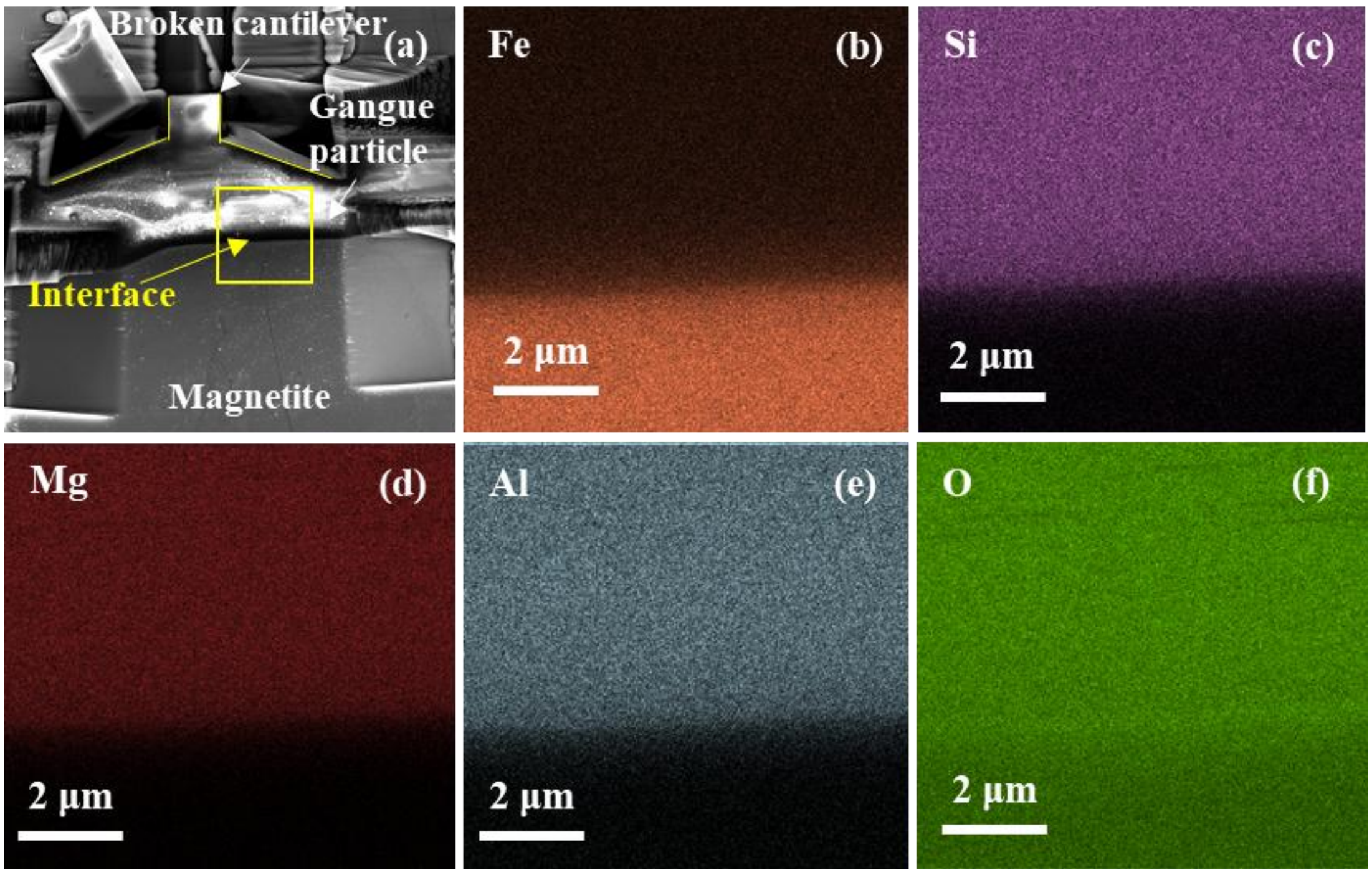

Fig. S1: (a) SEM image of gangue particle-magnetite interface shown with yellow arrow(b-f) EDS analysis of the gangue particle-magnetite showing spatial distribution of Fe, Si, Mg, Al and O.

EDS was conducted in the highlighted region in Fig. S2a (highlighted in yellow box) corresponding to the magnetite-gangue particle interface (highlighted with yellow →) using a Jeol® JSM-IT510 SEM equipped with EDAX EDS system. The distribution of Fe shows the magnetite region (Fig. S2 b). The opposite side of the interface shows presence of Si, Mg and Al in oxide indicating it to be a gangue particle.

## 3. Constraint factor and general yield strength

As highlighted in Section 2.3 in the manuscript, determining the size of the plastic zone ahead of the crack tip is essential for a valid $\boldsymbol{K_{IC}}$ measurement, as it is directly influenced by yield strength $\sigma_y$, as shown in Eq. 1. Since there is little apparent strain hardening as observed during micropillar compression of the three iron oxide phases (refer to Fig. S4 of Supplementary Information for micropillar compression results) it is reasonable to assume $\sigma_y \sim \sigma_f$ , where $\sigma_f$ is the flow stress.

To derive the representative flow stresses $\boldsymbol{\sigma_f}$ from the indentation data, it is first necessary to estimate the constraint factor, $C*$ which establishes the relationship between hardness and the equivalent flow stress, as proposed by Tabor in accordance with Eq. S1 [1,2]:

$$H = C^{*}\sigma_f \quad [S1]$$

Consequently, $C*$ for all three phases was determined from the Berkovich nanoindentation experiments, following the methodology proposed by Hay et al [2], by utilizing the loading and unloading stiffness at each point within the 50–300 nm displacement range, as plotted in Fig. S3a. Due to the self-similar nature of the Berkovich indenter, no variation in $C*$ was observed with indentation depth.

The vales of the $C*$ obtained from this analysis for the 3 iron oxide phases are reported in Fig. S3b. Based on the $C*$ and $H$, the flow stress ($\boldsymbol{\sigma_f}$) for hematite, magnetite and Wüstite was calculated to be 8.2±0.2, 3.7±0.1, 3.0±0.1 GPa respectively. It is important to note that in Eq. 1 the plastic zone size is calculated based on the yield strength ($\boldsymbol{\sigma_y}$), however the Eq. S1 here calculates the $\sigma_f$ which corresponds to flow stress at representative strain of ~8% in case of Berkovich indenter.

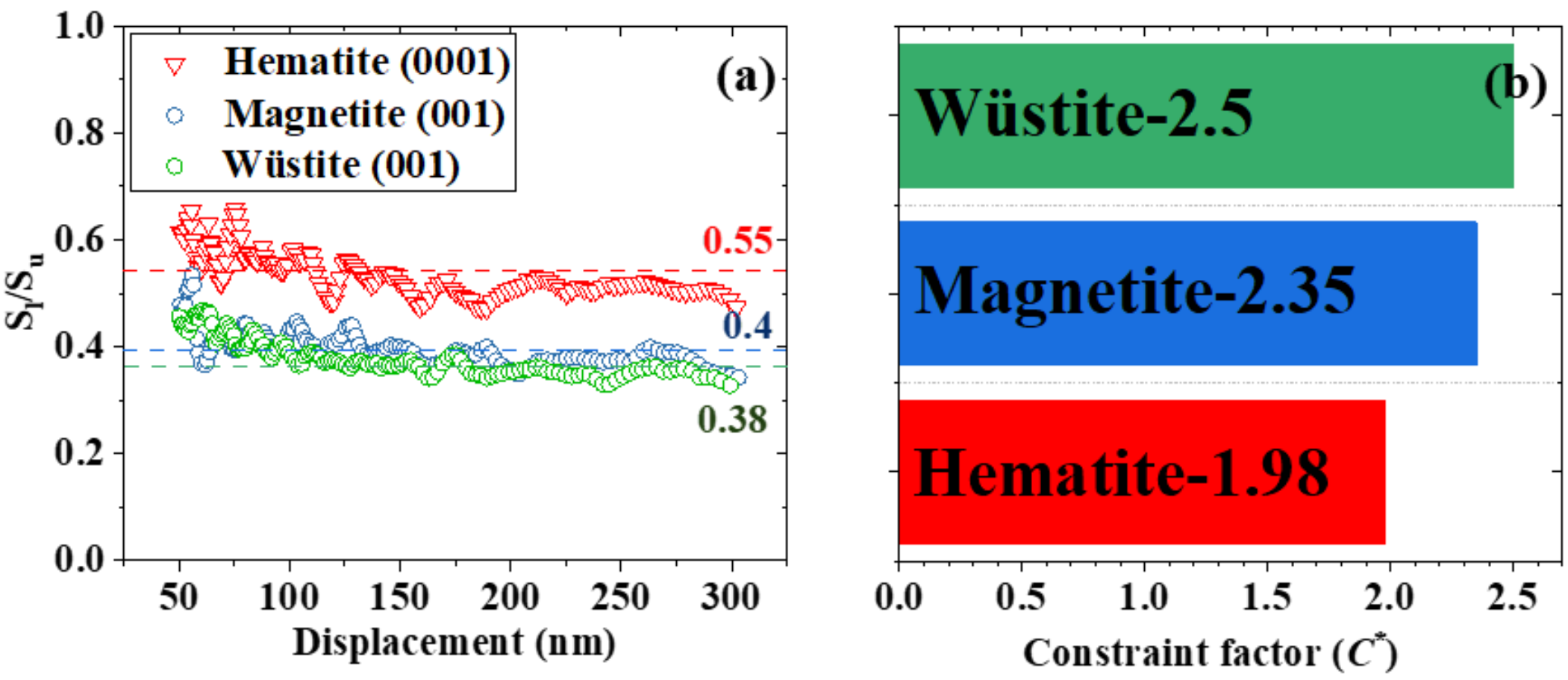

**Fig. S2**: (a) Variation of the ratio of loading to unloading stiffness ($S_l/S_u$) with respect to indentation depth. (b) Constraint factor ($\boldsymbol{C^*}$) measured from ($S_l/S_u$) for hematite, magnetite and Wüstite.

**4: Engineering stress vs Engineering strain curve for hematite, magnetite and Wüstite.**

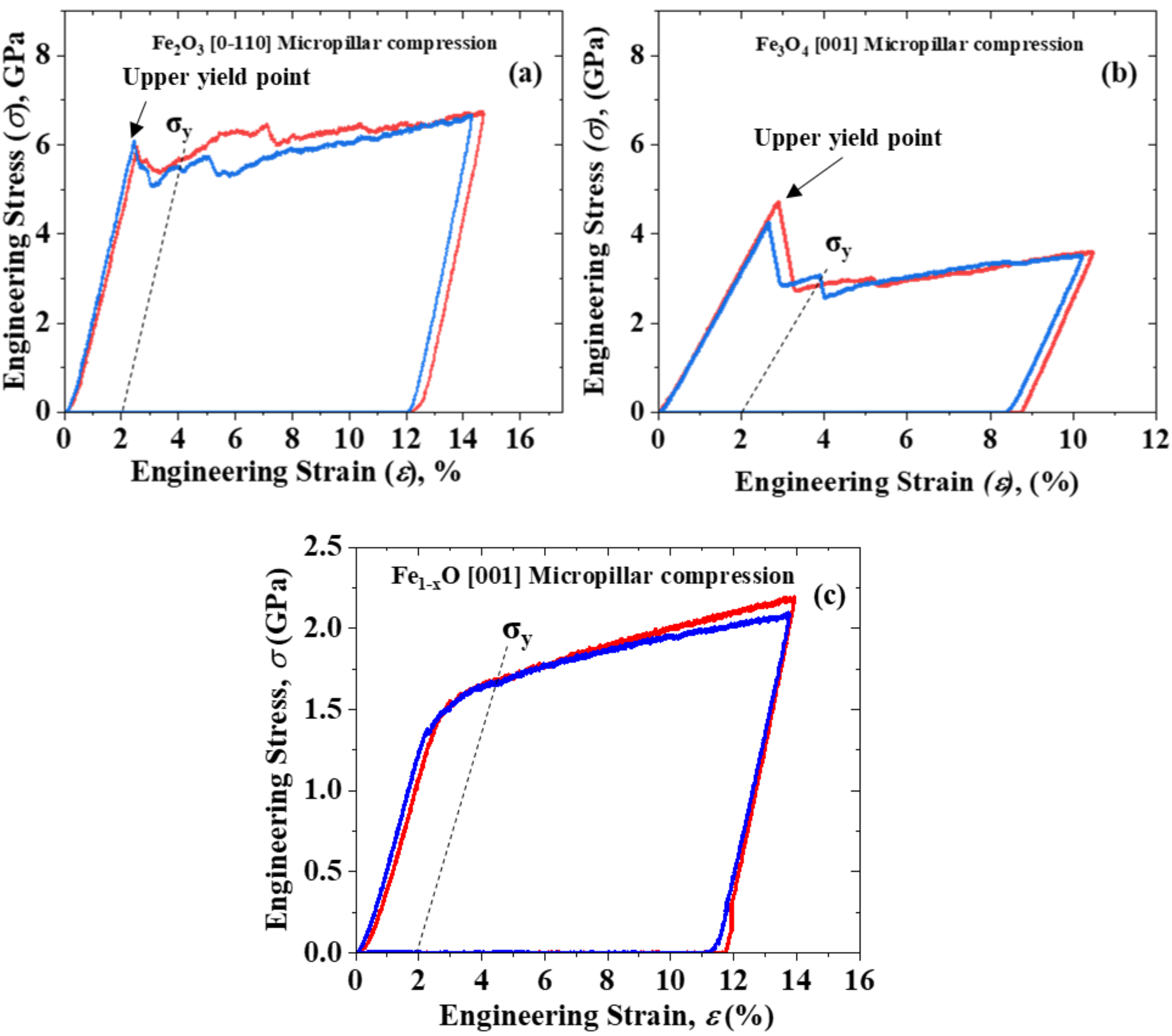

Fig. S3: Representative engineering stress *(σ)* vs engineering strain *(ε)* curve for hematite, magnetite and Wüstite.

The representative $\boldsymbol{\sigma}$ - $\boldsymbol{\varepsilon}$ curves from micropillar compression for hematite, magnetite and Wüstite shows little to no strain hardening. In the case of hematite and magnetite the initiation of plasticity is marked by a sudden stress drop (marked as "upper yield point") as marked in the Fig. S3 a,b. In the case of Wüstite, the transition from elastic to plastic was smooth. In all the 3 cases the yield strength ($\sigma_y$) is determined by 2% offset slope as shown in Fig. S3. In all the 3 oxides the plastic flow proceeds at near constant stress with little apparent strain hardening, hence its fair to assume $\sigma_f \sim \sigma_y$

**5: EBSD analysis showing orientation of low index oriented microcantilever**

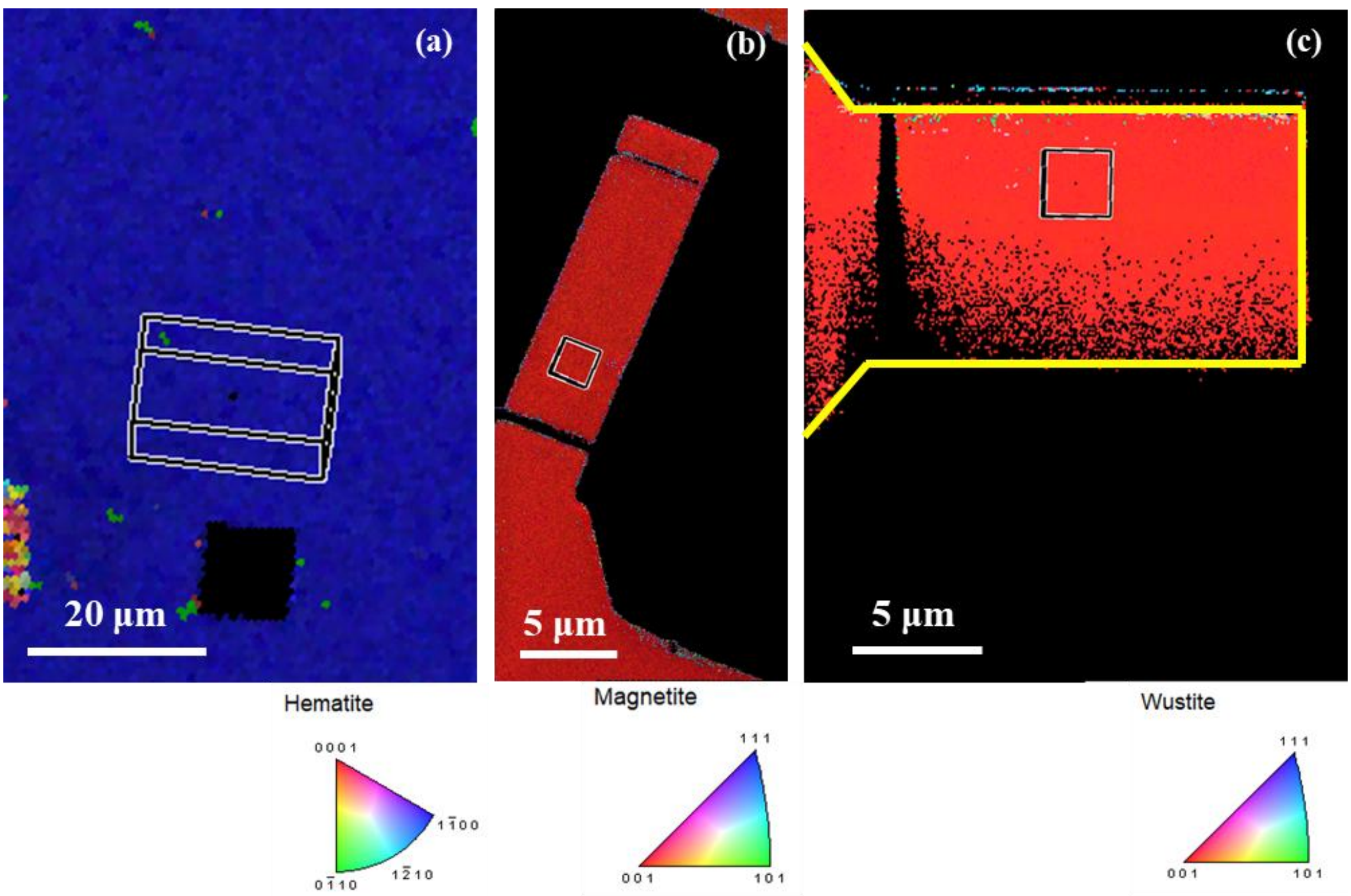

Fig. S4: EBSD showing the orientation of the low index oriented notch in (a) hematite, (b) magnetite and (c) Wüstite.

EBSD analysis was performed on the microcantilevers after the final milling step. In the case of hematite, resolvable Kikuchi patterns could not be obtained on the microcantilever or the surrounding region. This absence of signal may be attributed to redeposition or ion-beam-induced damage in areas exposed to $Xe^{+}$ irradiation. To establish a crystallographic reference, a FIB marker aligned with the notch of the microcantilever was fabricated using a very low ion current (30 pA). EBSD mapping was subsequently conducted using this marker as a reference, as illustrated in Fig. S4(a). The edge of the marker aligned with the notch was found to be parallel to the basal plane, thereby confirming the notch orientation. Fig. S4(b) and (c) correspond to EBSD results from magnetite and Wüstite microcantilevers, respectively. In both cases, the bridge notch is oriented along the (001) plane. For the Wüstite sample, the microcantilever was milled near the edge of the specimen, resulting in a slight rounding effect. This rounding caused the top surface of the microcantilever to deviate from planarity, leading to reduced EBSD signal quality near one edge of the microcantilever, as shown in Fig. S4(c).

## 6: Geometric details of low index microcantilevers of hematite, magnetite and Wüstite.

**Table: S1**

| Sample No | Cross section shape | L (in µm) | W (in µm) | B (in µm) | a (in µm) | $P_{max}$ (mN) | Fracture Toughness (MPa$\sqrt{m}$) |
|---|---|---|---|---|---|---|---|
| **Fe2O3 0001 C1** | Rectangular | 15.41 | 5.13 | 7.2 | 1.07 | 3.75 | 3.57 |
| **Fe2O3 0001 C2** | | 15.56 | 5.03 | 6.8 | 1.16 | 3.76 | 4.18 |
| **Fe2O3 0001 C3** | | 15.51 | 5.26 | 6.4 | 1.07 | 3.34 | 3.41 |
| **FeO 001 C1** | | 15.07 | 4.99 | 6.83 | 1.33 | 1.94 | 2.31 |
| **FeO 001 C2** | | 15.24 | 5.13 | 6.47 | 1.335 | 1.9 | 2.28 |
| **FeO 001 C3** | | 15.13 | 5.44 | 6.3 | 1.24 | 1.72 | 1.77 |
| **FeO 001 C4** | | 15.38 | 5.13 | 6.33 | 1.25 | 1.54 | 1.82 |

**Table: S2**

| Sample ID | Cross section shape | L (in µm) | W (in µm) | B (in µm) | h (in µm) | a (in µm) | $P_{max}$ [mN] | Fracture Toughness (MPa$\sqrt{m}$ |
|---|---|---|---|---|---|---|---|---|
| **Fe3O4 C1** | Pentagonal | 13.2 | 5.08 | 4.97 | 7.16 | 1.17 | 2.24 | 2.87 |
| **Fe3O4 C2** | | 13.37 | 5.47 | 4.26 | 6.68 | 1.12 | 2.02 | 2.65 |
| **Fe3O4 C3** | | 12.68 | 5.26 | 4.51 | 7.16 | 1.15 | 2.41 | 2.80 |

**7: EBSD analysis showing orientation of high index oriented microcantilever.**

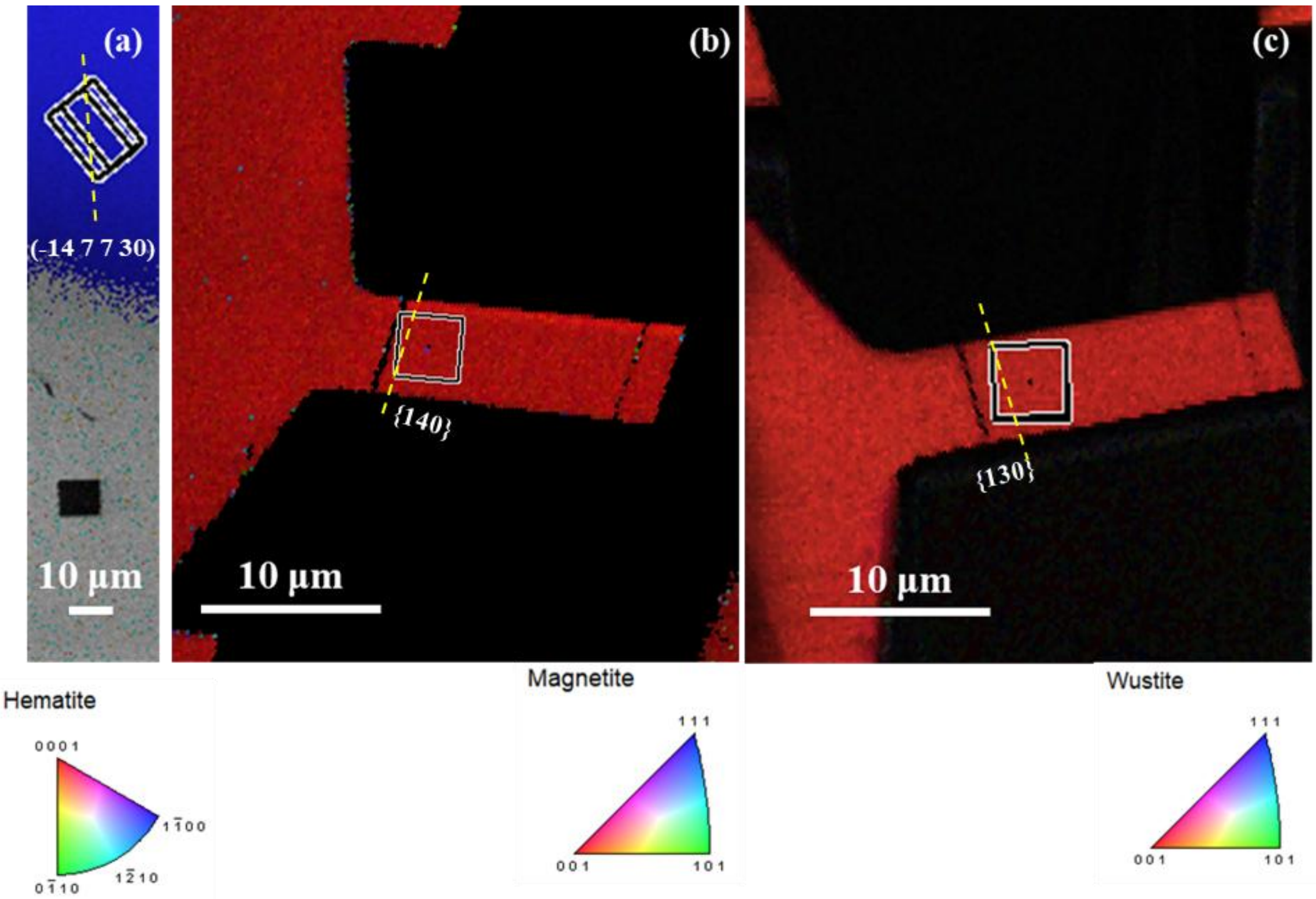

Fig. S5: EBSD showing the orientation of the high index-oriented notch in (a) hematite, (b) magnetite and (c) Wüstite.

EBSD was performed on the microcantilevers after the final milling. For the reasons mentioned in Sec. 4 direct EBSD on the hematite microcantilevers could not be performed hence reference FIB marker was used. For magnetite and Wüstite same protocol as Sec. 4 was followed.

## 8: Geometric details of high index microcantilevers of hematite, magnetite and Wüstite.

**Table: S3**

| Sample ID | Cross section shape | L (in µm) | W (in µm) | B (in µm) | a (in µm) | $P_{max}$ (mN) | Fracture Toughness (MPa$\sqrt{m}$) |
|---|---|---|---|---|---|---|---|
| Fe2O3 C1 | Rectangular | 14.36 | 5.28 | 7.75 | 1.05 | 3.98 | 3.05 |
| Fe2O3 C2 | | 14.51 | 5.3 | 7.72 | 1.02 | 4.45 | 3.38 |
| Fe2O3 C3 | | 14.54 | 5.32 | 8.76 | 1.08 | 5.62 | 3.86 |

**Table: S4**

| Sample ID | Cross section shape | L (in µm) | W (in µm) | B (in µm) | h (in µm) | a (in µm) | $P_{max}$ [mN] | Fracture Toughness (MPa$\sqrt{m}$ |
|---|---|---|---|---|---|---|---|---|
| Fe3O4 C1 | Pentagonal | 15.49 | 4.885 | 5.82 | 8.01 | 1.12 | 4.21 | 4.79 |
| Fe3O4 C2 | | 14.4 | 6.75 | 6.5 | 9.5 | 1.07 | 7.21 | 4.25 |
| Fe3O4 C3 | | 14.4 | 6.97 | 6.1 | 8.77 | 1.05 | 6.42 | 4.23 |
| FeO C1 | | 13.79 | 6.53 | 6.19 | 8.76 | 1.07 | 4.1 | 2.89 |
| FeO C2 | | 13.87 | 6.06 | 6.15 | 9.01 | 1.04 | 4.15 | 2.90 |
| FeO C3 | | 13.92 | 6.22 | 6.13 | 8.65 | 1.03 | 3.75 | 2.77 |

## 9: Measurement of flaw dimensions from un-notched cantilevers

To check the validity of linear elastic fracture mechanics (LEFM) un-notched simple cantilever of hematite and Wüstite, oriented same as high index micro cantilevers was fractured under the same conditions. Fig. S6 a,b shows the load vs displacement for un-notched cantilevers of hematite and magnetite. The maximum load corresponds to the point of fracture which is used to calculated the fracture stress based on Eq. S2 and S3. For hematite the microcantilevers have rectangular cross section hence Eq. S2 is used

$$\sigma_{max} = \frac{6FL}{bh^2} \quad \text{S2}$$

Where F is the fracture load, L is the distance between the indenter loading point and fixed end, b is the width and h is the thickness of the microcantilever, as described in Fig 1a.

For pentagonal cross section the general formula for maximum stress is given by Eq. S3:

$$\sigma_{max} = \frac{FLy}{I} \quad \text{S3}$$

Where *y* is the distance of the top fiber from the neutral plane which is calculated using Eq. S4 and *I* is the moment of inertia of the pentagonal beams given by Eq. S5

$$y = \frac{h^2+hb+b^2}{3(h+b)} \quad \text{S4}$$

$$I = \frac{1}{3}wy^2 + \frac{1}{3}w(b-y)^3 + \frac{1}{2}w(h-b)(b-y)^2 + \frac{1}{3}w(h-b)^2(b-c) + \frac{1}{12}w(h-b)^3 \quad \text{S5}$$

The dimensions w, b and h has been described in Fig. 1b.

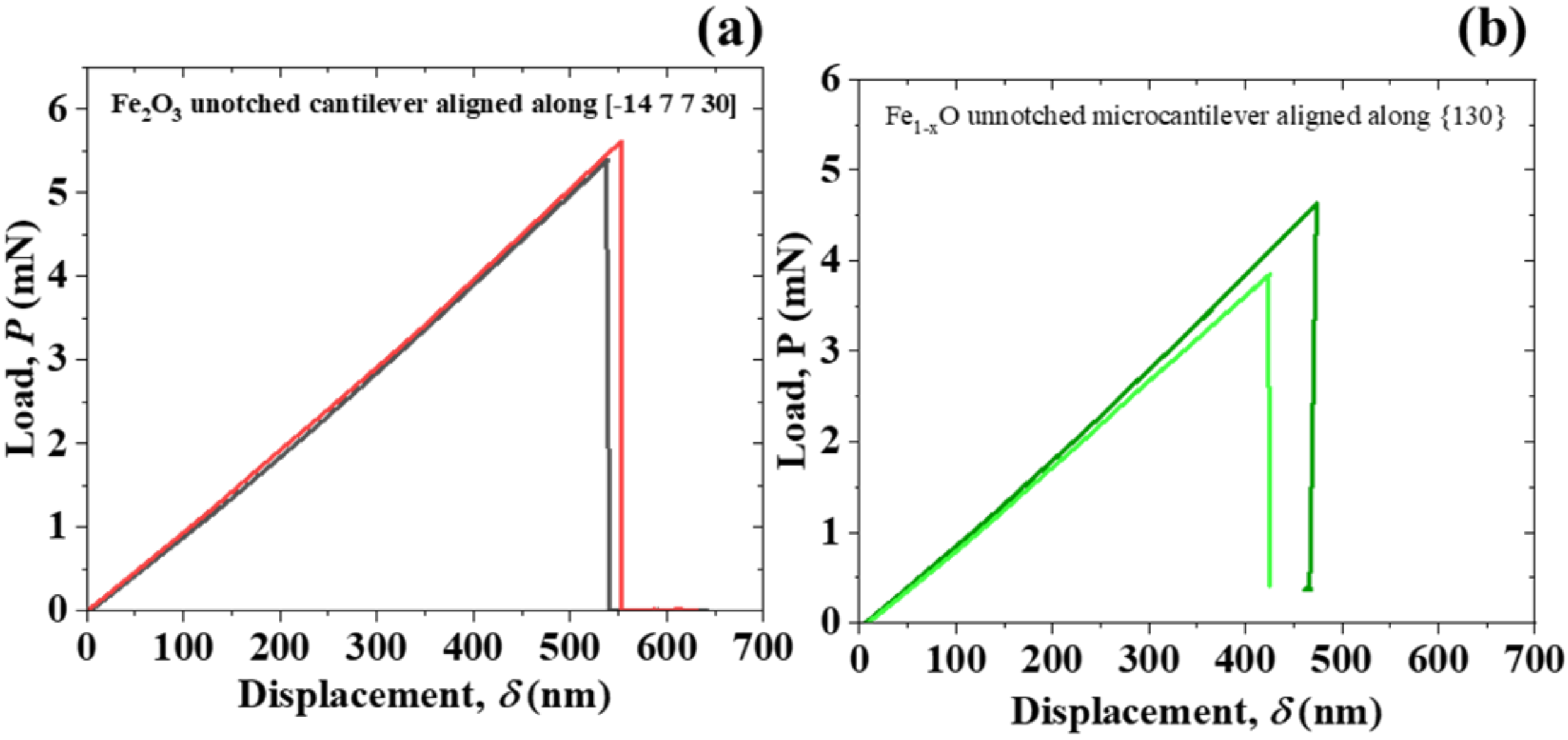

Fig. S6: Load–displacement curves for fracture of unnotched microcantilevers of hematite (a) and Wüstite (b) oriented along the (-14 7 7 30) and {130} planes respectively.

The load-displacement curve for unnotched hematite cantilever is shown in Fig S6a. The cantilever is aligned along the same orientation as high index orientation (-14 7 7 30). The maximum stress on the top fiber of the cantilever was measured from Eq. S2. The fracture toughness is known for this orientation from Sec. 3.4.2. The relationship between the flaw size, maximum stress and fracture toughness in LEFM is given by Eq. S6

$$\sigma_{max} = \frac{K_{IC}}{Y\sqrt{\pi c}} \quad \text{S6}$$

Where $K_{IC}$ is the measured mode 1 fracture toughness for that given orientation, $c$ is the flaw dimension and $Y$ is the geometry factor. For a surface flaw under tensile loading, $Y$ is 1.12 [3]. Since $K_{IC}$ and $\sigma_{max}$ is known the effective flaw size '$c$' can be back calculated from Eq. S6. For hematite cantilever aligned along the (-14 7 7 30) the maximum allowable flaw size was calculated to be 0.82 µm and 0.8µm for the two cantilevers tested. The critical flaw size estimated from unnotched strength (~0.8 µm) is in close agreement with the actual notch depth (~1 µm), Similarly for Wüstite the representative load vs displacement curves for unnotched microcantilevers aligned along {130} plane is shown in Fig. S6b. Using equations S3, S4, and S5 the effective flaw size was calculated to be 0.82 µm and 0.75 µm which is in close agreement to the actual notch depth (~1µm.). The close agreement between the effective flaw size and actual notch depth for both hematite and magnetite indicates that the fracture is governed by LEFM.

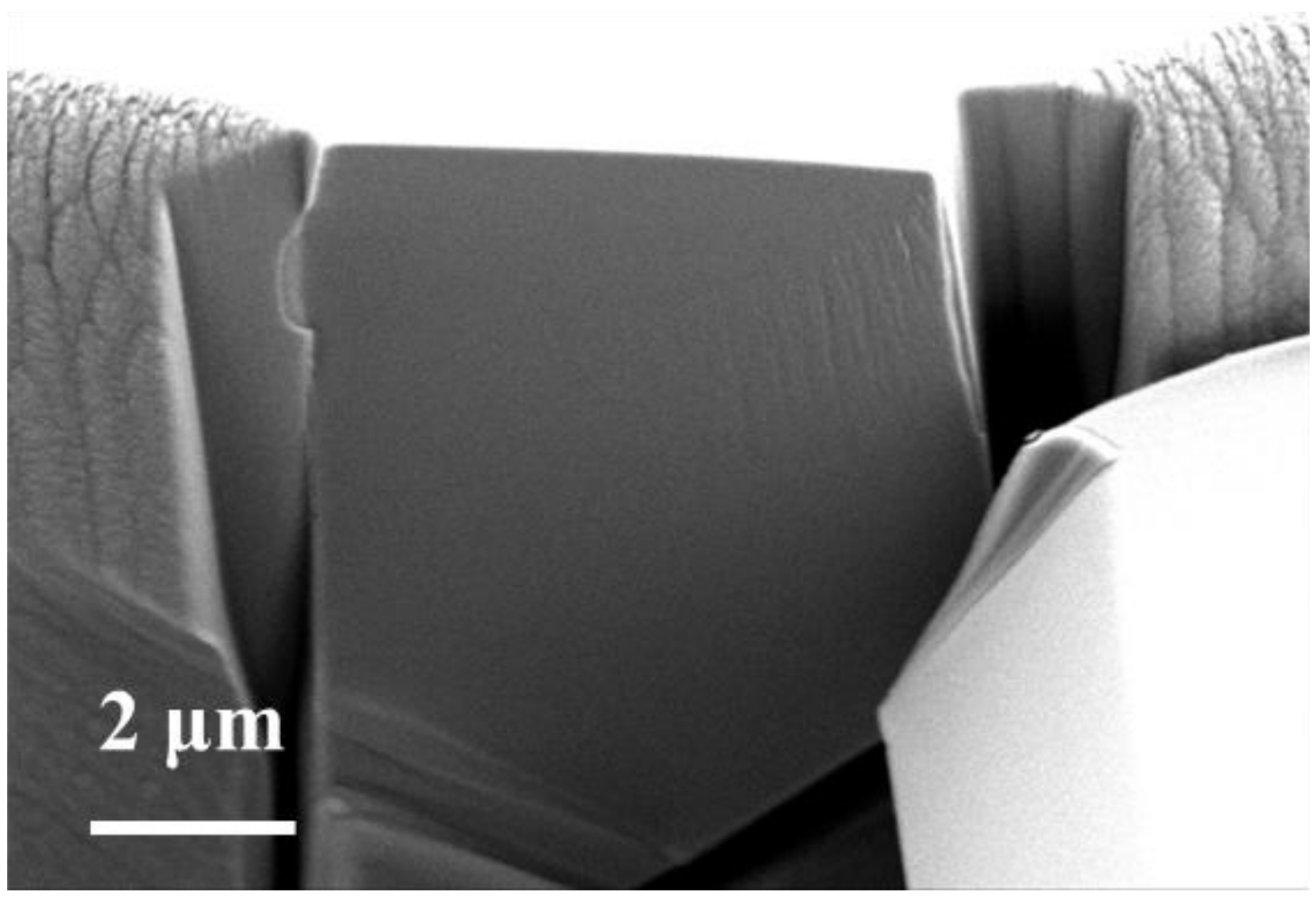

Fig. S7: SEM image of the fracture surface corresponding to an unnotched Wüstite microcantilever aligned along {130} plane.

Fig. S7 shows the fracture surface of an unnotched Wüstite microcantilever aligned along the {130} plane. The fracture surface exhibits a clean cleavage failure along the {130} plane without any observable ratchet marks. This indicates that the ratchets observed in notched {130} oriented Wüstite could be from notch tip inhomogeneity as described in Sec. 4.2.